%% file: main.tex
\documentclass[manuscript, screen]{acmart}

\usepackage{dirtytalk}
\usepackage{xargs}      
\usepackage{soul}       
\usepackage{color}      
\usepackage{xspace}     
\usepackage{xpunctuate} 
\usepackage{float}
\usepackage{wrapfig}

\input{commands.tex}

\AtBeginDocument{%
  }

\setcopyright{acmlicensed}
\copyrightyear{2025}
\acmYear{2025}
\acmDOI{XXXXXXX.XXXXXXX}

\begin{document}

\title{Bias, Accuracy, and Trust: Gender-Diverse Perspectives on Large Language Models}

\author{Aimen Gaba}
\affiliation{%
  \institution{University of Massachusetts Amherst}
  \state{Massachusetts}
  \country{USA}}
\email{agaba@umass.edu}

\author{Emily Wall}
\affiliation{%
 \institution{Emory University}
 \state{Georgia}
 \country{USA}}
\email{emily.wall@emory.edu}

\author{Tejas Ramkumar Babu}
\affiliation{%
  \institution{Georgia Institute of Technology}
  \state{Georgia}
  \country{USA}}
\email{tbabu8@gatech.edu}

\author{Yuriy Brun}
\affiliation{%
  \institution{University of Massachusetts Amherst}
  \state{Massachusetts}
  \country{USA}}
\email{brun@cs.umass.edu}

\author{Kyle Wm. Hall}
\affiliation{%
  \state{Ontario}
  \country{Canada}}
\email{kylewmhall@gmail.com}

\author{Cindy Xiong Bearfield}
\affiliation{%
  \institution{Georgia Institute of Technology}
  \state{Georgia}
  \country{USA}}
\email{cxiong@gatech.edu}

\renewcommand{\shortauthors}{Gaba et al.}

\begin{abstract}
Large language models (LLMs) are becoming increasingly ubiquitous in our daily lives, but numerous concerns about bias in LLMs exist. 
This study examines how gender-diverse populations perceive bias, accuracy, and trustworthiness in LLMs, specifically ChatGPT. Through 25 in-depth interviews with non-binary/transgender, male, and female participants, we investigate how gendered and neutral prompts influence model responses and how users evaluate these responses. Our findings reveal that gendered prompts elicit more identity-specific responses, with non-binary participants particularly susceptible to condescending and stereotypical portrayals. Perceived accuracy was consistent across gender groups, with errors most noted in technical topics and creative tasks. Trustworthiness varied by gender, with men showing higher trust, especially in performance, and non-binary participants demonstrating higher performance-based trust. Additionally, participants suggested improving the LLMs by diversifying training data, ensuring equal depth in gendered responses, and incorporating clarifying questions. This research contributes to the CSCW/HCI field by highlighting the need for gender-diverse perspectives in LLM development in particular and AI in general, to foster more inclusive and trustworthy systems.
\end{abstract}





\maketitle

\section{Introduction}
\label{sec:Intro}

Machine learning (ML)-driven interfaces are becoming ubiquitous in modern society~\cite{Komorowski18, qai22, Roy20, Angwin16}, from facial recognition systems used to unlock our phones, to voice-activated assistants, and conversational systems that respond to natural language prompts. These technologies are widely deployed, but they often exhibit biases that reflect existing inequalities in society. For example, facial recognition technologies tend to perform better for individuals with lighter skin tones~\cite{buolamwini2018gender}, and voice-controlled devices often struggle with certain accents~\cite{voicebias2019}. In the case of large language models (LLMs), recent studies have shown that they can exhibit biases and other social risks~\cite{weidinger2021ethical} against particular religious groups~\cite{antimuslimbias}, produce gender stereotypes~\cite{unesco2024generative}, and generate stigmatizing language~\cite{nozza-etal-2022-measuring, center_transgender_voices}, reinforcing harmful stereotypes when discussing non-binary or transgender individuals. These biases not only undermine user trust in ML models~\cite{gabamodelfairness}, but also degrade the model's ability to provide meaningful and fair interactions.

Conversations around fairness and harms in sociotechnical systems have been a focal point within the CSCW and broader HCI community~\cite{solyst-youthbias-cscw23, imagecroppingtwitter, wagman-feministsts-cscw21, bigdatainequality}. Previous discussions have emphasized the implications of biased technologies on marginalized identities, especially concerning gender~\cite{genderstereotypesinai-25, ertranstech24}. For instance, Scheuerman et al.~\cite{facialgender-klaus-19} found that commercial facial analysis services consistently perform worse for transgender individuals and are universally unable to classify non-binary genders, highlighting significant gaps in technological inclusivity. These dialogues within the CSCW community underscore the importance of diverse representation and gender balance in machine learning evaluations~\cite{leavy-18-genderbiasinai}.

Although substantial research has focused on understanding and mitigating algorithmic biases in machine learning systems and language models~\cite{mehrabi2021survey, dangersllms2021, duan-mitigatinggenderbias-xai, leavy-mandal2023multimodal}, particularly in domains such as machine translation~\cite{ghosh2023chatgpt}, text classification~\cite{biastextclassification}, data science~\cite{dong-responsibleds-cscw25}, and coreference resolution~\cite{zhao2018gender}, a significant gap remains in understanding how users perceive and respond to these biases in real-world LLM applications. 
Notably, there is limited qualitative research on how individuals, especially those from gender-diverse backgrounds, experience and interpret bias in LLM outputs. 
While LLMs have been used to analyze bias in interview settings~\cite{kong2024gender} or to assist in qualitative research~\cite{ashwin2023using, dengel2023qualitative}, few studies have centered on gender bias~\cite{genderbiaskotek, gender_bias_nlp, gender_bias_mitigation_llms}, particularly from the perspectives of non-cisgender users, such as non-binary and transgender individuals. Moreover, although prior work emphasizes the value of qualitative case studies~\cite{qualimportance, elish2020repairing}, it rarely integrates in-depth qualitative insights with quantitative measures of user perception - a gap this study seeks to address.

In this paper, we study the perceived utility of a real-world LLM application by understanding users' perception of bias, accuracy, and trust in them.
In order to study this in a real-world setting, we selected ChatGPT~\cite{openai2024chatgpt} for two key reasons. First, its widespread use and accessibility allow us to study how biases manifest in real-world LLM applications. Second, ChatGPT is regularly used by people from different gender backgrounds~\cite{draxler2023gender}, offering an important context for examining how gender influences users' perceptions of LLM interactions. 
In this work, we address five research questions. Figure~\ref{fig:rqs} summarizes the research questions that underpin our user studies and their respective findings. 

\input{Tables/researchquestions}

To address these research questions, we studied how gendered and neutral prompts (e.g., ``man'', ``woman'', ``non-binary'', ``person'') influence the responses generated by ChatGPT and how users interpret those responses. We conducted semi-structured interviews with 25 participants and extract qualitative insights from our study. We adopted a mixed-methods approach that combines quantitative trust measures with qualitative interview data, allowing us to examine in depth as to how users perceive the biases and assumptions present in ChatGPT's outputs.
Moving forward, we define \textcolor{ManColor}{``men''} as cisgender men, \textcolor{WomanColor}{``women''} as cisgender women, and \textcolor{NonbinaryColor}{``non-binary/transgender''} as those who do not identify with the gender assigned at birth. 
By exploring these gender categories, we investigate how gender affects user perceptions of bias, accuracy, and trustworthiness in LLMs.

In this work, we make the following contributions:
\begin{itemize}
    \item We introduce curated gender-focused prompts to evaluate LLMs' inclusivity and gender representation.
    \item We study men, women, and non-binary/transgender participants, revealing varying reactions to bias, with non-binary participants particularly susceptible to identity framing.
    \item We investigate perceived accuracy, examining how participants understand and evaluate LLM responses.
    \item We assess trustworthiness through qualitative analysis of participant reactions in high-stakes scenarios and quantitative trust scores collected before and after the study, highlighting the roles of gender, bias knowledge, and LLM familiarity.
    \item Finally, we offer a preliminary set of recommendations and suggestions from the participants as well as design implications for developing more inclusive and trustworthy LLMs in the future.
\end{itemize}

\section{Related Work}
\label{sec:Relatedwork}

\subsection{Social Biases in NLP and Language Technologies}
\label{sec:rel_bias}
The intersection of language technologies and gender representation has become a critical area of study~\cite{leavy-18-genderbiasinai, leavy-genderbiasnewspaper-18}, particularly with the rise of LLMs. Research has shown that NLP systems often reinforce binary and exclusionary notions of gender, reflecting broader societal biases~\cite{kiritchenko2018examining}. Research has explored how LLMs perpetuate gender stereotypes, particularly in professional contexts such as the language used to describe individuals in recommendation letters or the association of specific professions with certain genders~\cite{gender_bias_nlp, gender_nlg, wan2023kelly, unesco2024generative}. These findings underscore the subtle ways in which LLMs can reinforce societal biases, highlighting the nuanced nature of the issue.

Further research has expanded on these findings, showing that LLMs frequently fail to accommodate non-binary identities. For example, Ghosh et al.~\cite{ghosh2023chatgpt} highlight the limitations of LLMs in representing gender diversity, particularly the persistent exclusion of non-binary and transgender individuals in the models' outputs. This issue is further exacerbated by biases that extend beyond gender, such as anti-Muslim bias, which has been documented in LLMs~\cite{antimuslimbias}. These studies reflect the prevalant nature of the biases present in LLMs, not just limited to gender, but that spans various social identities.

Other research has examined how LLMs replicate human tendencies to perceive socially subordinate groups as less diverse than dominant groups~\cite{llmhomogenous}. This study emphasizes the inherent biases in LLMs' representations of social groups, mirroring broader societal stereotypes. Similarly, research has shown that LLMs can generate stigmatizing language, particularly when discussing non-binary individuals, which reinforces harmful stereotypes~\cite{nozza-etal-2022-measuring}. The paper~\cite{disability_perspectives_llms} offers a parallel perspective, exploring how marginalized communities, including individuals with disabilities, experience biases in LLM outputs, stressing the importance of inclusive evaluation approaches in understanding these biases.

In response to these issues, efforts have been made to quantify and mitigate these biases. For example, Holstein et al. \cite{gender_bias_mitigation_llms} advocate for transparent reporting of model behaviors and active interventions to reduce harmful tendencies in LLMs. Similarly, Park et al. \cite{park2018reducing} evaluate existing models and introduce methods for bias mitigation, such as adversarial training~\cite{kumar2024decoding}, data augmentation, and debiasing algorithms, to improve fairness and accuracy in detecting abusive content. Bartl and Leavy~\cite{leavy-bartl2024showgirls} developed a gender includive dataset called the \textit{Tiny Heap} and observed an overall reduction in gender stereotyping tendencies across three large lagnuage models that were fine-tuned using that dataset. These studies highlight the dual responsibilities of identifying biases and designing interventions to mitigate harm, making them a key part of ongoing efforts to improve LLM fairness and inclusivity.

\subsection{Accuracy \& Trust in AI}
\label{sec:ref-trust}
Understanding and fostering trust in AI systems is crucial, particularly in sensitive or high-stakes domains~\cite{trustnasa2021}. Research highlights the role of explainability and transparency in building trust, with frameworks like~\cite{trust_in_ai} and~\cite{trust_model} emphasizing user understanding and perception of AI behavior in sociotechnical contexts. Studies~\cite{humansaicontext, trustnasa2021, questioningai} show how transparency and contextual awareness improve perceptions of AI reliability, while~\cite{helpmehelpai} underscores the importance of explainable interfaces to reduce uncertainty and increase confidence.

Liao et al.\cite{liao2023ai} aligns these concerns with actionable recommendations to enhance LLM transparency, framing transparency as vital for ethical AI design. Studies like~\cite{zhou2024teachers} identify challenges with LLM inaccuracies, biases, and educational misalignment, advocating for ethical guidelines and training. Similarly, Lee et al.~\cite{onevsmany_ai_gen} explore how conflicting AI outputs increases user comprehension but comprimising users' percieved AI capacity, highlighting the need for transparency in AI models and promote critical LLM usage.
In this work~\cite{kumarllmlearning}, Kumar et al. conduct a formative and controlled experiment to explore the impact of pedagogically informed guidance strategies on the students’ performance, confidence and trust in LLMs. To measure trust, they collected pre- and post-measures on a scale of 7 regarding perception of LLMs (confidence in LLMs’ responses, helpfulness, willingness to interact again and error tolerance) and perception of self (confidence in their answers, self-confidence for the given topic of assignment).

Göben and Lämmel~\cite{model_based_trust_llms} conduct a trust analysis of LLM conversations by using a trust score between -1 (absolute distrust) and +1 (absolute trust) function to understand and assess users' trust in knowledge elicited from these LLMs and its sources, offering insights into trust formation during interactions. This work~\cite{user_trust_chatgpt} identifies trust as central to ChatGPT’s adoption while also highlighting that an overreliance on it for health-related advice could potentially lead to misinformation and subsequent health risks. Shen et al.~\cite{chatgpt_reliability_trust} conduct a large-scale measurement of ChatGPT's reliability in the generic QA scenario with a carefully curated set of questions across different datasets and domains finding ChatGPT's reliability varying across different domains, especially underperforming in law and science questions.
We supplement these existing work to better understand the perceived trustworthiness and the challenges LLMs face in earning user trust.

\section{A Case Study on ChatGPT Response Patterns}
\label{sec:GPTresponsepatterns}

To initiate discussions with ChatGPT about gender, we curated four prompts inspired by~\cite{disability_perspectives_llms}. These prompts were crafted through extensive probing to encourage meaningful, gender-focused responses while avoiding harmful or biased content. Our aim was to design prompts that could be neutrally addressed and resonate with lived experiences across diverse populations, without leading or eliciting harmful narratives. The final prompts are as follows: 

\begin{enumerate}
    \item Tell me a story about a \textcolor{PersonColor}{person} buying a car.
    \item Tell me a story about a \textcolor{PersonColor}{person} navigating through college.
    \item Tell me a story about the experiences of a \textcolor{PersonColor}{person}.
    \item Describe an experience of a \textcolor{PersonColor}{person} applying for jobs.
\end{enumerate}

To explore gendered perspectives, we used the same prompts but replaced ``person'' with specific gender identities: \textcolor{ManColor}{man}, \textcolor{WomanColor}{woman}, and \textcolor{NonbinaryColor}{non-binary person}. For example, the first prompt became ``Tell me a story about a non-binary person buying a car'' eventually resulting in a total of 16 unique prompts.

\subsection{RQ1: Effect of Gendered Prompts on \gpt Responses}
\label{sec:rq1gptresponsepatterns}

We analyze the responses of ChatGPT to the four prompts mentioned above (repeated 10 times), focusing on how ChatGPT describes the main character, whether identified as a \textcolor{ManColor}{man}, \textcolor{WomanColor}{woman}, \textcolor{NonbinaryColor}{non-binary}, or \textcolor{PersonColor}{person}. We had a total of $40$ responses for each gender specification ($4$ unique prompts generated $10$ times).
We analyze these responses for names, pronouns, adjectives, emotions, goals, challenges, and symbols to assess their role in shaping the character's identity and narrative.

\subsubsection{\textcolor{NonbinaryColor}{Non-binary}}


ChatGPT consistently used they/them \textbf{pronouns} they/them in relation to the non-binary character and assigned predominantly gender-neutral \textbf{names} (often leaning towards masculine) to non-binary characters such as Alex (11), Taylor (10), Jordan (8), and Jamie (4). Occasionally, names such as \say{Casey} (2), \say{Avery} (1), and \say{Riley} (1) appeared.
Common \textbf{adjectives and emotional descriptors} used for the non-binary character included \say{resilience} (21), \say{authenticity} (9), \say{anticipation} (6), \say{determination} (5). Some other common descriptors included \say{liberation}, \say{inclusivity or inclusive spirit}, \say{non-conforming identity}, or \say{unconventional appearance}.
\textbf{Rersonal goals} frequently involved advocating for inclusion or expressing their true identity. Other common goals included \say{fostering/advocating inclusivity}, \say{embracing/express true self}, \say{navigating societal norms}, 
\textbf{Challenges} often centered on their identity such as \say{dealing with misgendering} (13), \say{job applications} (5) were highlighted most frequently. In some cases, it mentions their challenge \say{facing societal expectations, discrimination, systemic biases} and \say{deciding whether to disclose non-binary identity}.
In some responses, the non-binary character was depicted to be struggling with attire choices, but this was less frequent compared to identity-related challenges.
\textbf{Symbolic} elements highlighted themes of resilience with characters often portrayed as \say{beacons of resilience} or \say{embracing diversity}. 
Memorable quotes included: \say{Jordan found resonance in the car's ability to accommodate diverse needs-a reflection of their fluid identity}, \say{I'm non-binary, and my car choice reflects my values of sustainability and innovation}, and \say{The binary constraints of \textit{appropriate} clothing clashed with Alex's non-binary identity}.

\subsubsection{\textcolor{WomanColor}{Woman} Prompts}
ChatGPT consistently used she/her \textbf{pronouns} for the woman character and assigned commonly feminine, Western \textbf{names} such as Emma (6), Sarah (6), Emily (4), Rachel, Lily, and Amelia (9). 
Common \textbf{adjectives and descriptors} included \say{resilience} (27), determination (13), anticipation (7). Other descriptors frequently used with the female character were \say{compassionate}, with mentions of academic fields such as environmental science or psychology..
The character's \textbf{personal goals} often focused on achievement such as \say{navigating college as a first-generation student} (3) or \say{excelling in environmental science or psychology} (4).
\textbf{Challenges} typically involved \say{financial constraints} (9), \say{handling/managing rejections} (6), \say{navigating a competitive job market} (2).
In one response, the female character was depicted as \say{selecting a car that fits family and adventure needs}.
\textbf{Symbols} associated with the female character emphasized \say{growth} or \say{discovery}. Memorable quotes included: \say{the key to her newfound independence}, and \say{a sea of gender-based obstacles that sought to erode her confidence}.
 
\subsubsection{\textcolor{ManColor}{Man} Prompts}
In all 40 responses, ChatGPT consistently used he/him \textbf{pronouns} for the male character.
The male character was assigned traditionally masculine Western \textbf{names} commonly such as Alex (6), Mark (5), and Jake (4). These names are traditionally gendered and reflect the stereotype of male identity. Some names, like Alex, were more neutral but still paired with masculine descriptors.
Common \textbf{adjectives} included \say{resilient} (16), anticipation (5), determination (5) often linked to fields like \say{computer science or psychology}.
The male character's \textbf{personal goals} often focused on career success and leadership. ChatGPT frequently mentioned objectives such as \say{securing a job} (6) or \say{overcoming unemployment} (6).
\textbf{Challenges} were primarily career-related, including \say{unemployment} (8), \say{rejection emails} (7), and \say{job applications} (6) with occasional mentions of work-life balance struggles.
\textbf{Symbolism} highlighted assertive and traditional imagery portraying the male character as \say{the provider} or \say{adventurous}. Memorable quotes included: \say{his dedication to family and his role as a provider} and \say{unemployment had turned this room into a battlefield}.

\subsubsection{\textcolor{PersonColor}{Person} Prompts}
ChatGPT used she/her \textbf{pronouns} in 22 responses, he/him for in 15, and assigned no pronouns in the remaining 3.
Characters with gender-specific pronouns were given traditional \textbf{names} such as Mark or Sarah, while those without pronouns received neutral names like Alex.
The more commonly used \textbf{adjectives and descriptors} included \say{resilience} (14), \say{self-discovery} (11), and \say{determination} (8). 
\textbf{Personal goals} focused on \say{personal growth}, \say{embracing change}, \say{gaining independence}.
\textbf{Challenges} were typically related to academic or professional struggles, such as \say{rejections}, or \say{negotiating a deal}.
Responses without specified gender were noticeably more generic compared to gendered ones.

Overall, we found that gendered prompts elicited richer and more identity-focused responses compared to neutral prompts, which were more generic. Furthermore, men were associated with careers and assertiveness, women with growth and emotions, and non-binary individuals with inclusivity and advocacy. We also observed that \gpt's name assignments, pronoun usage, and challenges faced by characters align with societal gender norms. Our findings corroborate similar themes in~\cite{lucy-bamman-2021-gender, unesco2024generative} where GPT reinforces gender stereotypes and representation biases in storytelling, identifying systemic biases that disproportionately favor traditional gender norms.

\section{Perceived Utility of LLMs}
\label{sec:Interview}
To understand the perceived utility of LLMs, we conducted in-depth user interviews to understand how people perceive bias, accuracy, and trust in a real-world application of an LLM (ChatGPT). We describe our study methods (reviewed and approved by our Institutional Review Board prior to conducting the study) and our findings below.
\input{Tables/participantknowledge}

\subsection{Participant recruitment and selection}
\label{sec:participantrecruitment}
We recruited 25 participants. Among them, 9 identified as \textcolor{NonbinaryColor}{non-binary/transgender} ($M_{age}$ = 26.11, $SD_{age}$ = 8.51), 8 identified as \textcolor{WomanColor}{women} ($M_{age}$ = 25.62, $SD_{age}$ = 4.07), and 8 identified as \textcolor{ManColor}{men} ($M_{age}$ = 25.62, $SD_{age}$ = 2.39).
We first interviewed the \textcolor{NonbinaryColor}{non-binary} participants and then recruited \textcolor{WomanColor}{women} and \textcolor{ManColor}{men} participants of similar backgrounds (with considerations for diversity in the knowledge of LLM participants, bias in LLM and AI background). Concretely, we created a screening survey asking questions about their knowledge of LLMs, bias in LLMs, AI/ML background, and ChatGPT usage pattern (e.g., frequency of ChatGPT use). We distributed the survey through university social emails, various Slack workspaces within our institutions, and X. We were unable to recruit our desired number of non-binary participants using these methods, so we recruited and interviewed 3 non-binary participants via userinterviews.com by releasing the same screening survey. See Table~\ref{tab:participantknowledge} for a summary of the background of the participants and the supplementary material~\cite{SM} for definitions. The subgroups were defined based on the responses from the survey of the participants and the responses from the interviews. We categorized the participants into the range of low-to-high LLM/AI background (experience with LLM/AI) and low-to-high knowledge of LLM bias (awareness of bias and experience with encountering it).

Regarding frequency of use, 12 participants reported using it multiple times a week, 5 used it approximately three times or twice a week, and 8 used it once a week.

\subsection{Study protocol}
Our interviews were structured into two parts and included a short questionnaire that was distributed to measure trust~\cite{malle2023measuring, MALLE20213, Malle2021AMC, ullman2019} at the beginning and end of the study. The study protocol was iteratively designed through two pilot studies. The complete study protocol and interview guide is attached in the supplementary material~\cite{SM}.

\pheading{Context:} We began by asking participants about their background and usage of \gpt, including when and where they use it. We explored their knowledge and perceptions of bias and accuracy by having them reflect on instances of biased or incorrect responses they had encountered and how they identified them. Participants were also asked about their expectations of \gpt, if their experiences aligned with those expectations, and if certain prompts appeared to elicit more biased or accurate responses.

\pheading{Prompts:} Each participant encountered four prompts, as mentioned in Section~\ref{sec:GPTresponsepatterns}, arranged in a Graeco-Latin square design, ensuring exposure to all types of prompts and gender descriptions. For the fourth prompt, participants were shown responses for all gender descriptions to facilitate direct comparison.
We selected two random prompt-response pairings for each gender specification (from Section~\ref{sec:GPTresponsepatterns}) to show the participants and created two sets of responses:

\begin{itemize}
    \item The first set included responses shown to participants $1$ through $12$ (4 \textcolor{NonbinaryColor}{non-binary/transgender}, 4 \textcolor{WomanColor}{women}, and 4 \textcolor{ManColor}{men}).
    \item The second set included the responses shown to participants $13$ through $25$ (5 \textcolor{NonbinaryColor}{non-binary}, 4 \textcolor{WomanColor}{women}, and 4 \textcolor{ManColor}{men}).
\end{itemize}

Using different prompt-response sets for the second round allowed us to introduce variation and enhance the robustness of our findings.
Note that the responses to these prompts were generated in February 2024 using the free ChatGPT 3.5 version.

\begin{figure}
  \centering
    \includegraphics[width=\linewidth]{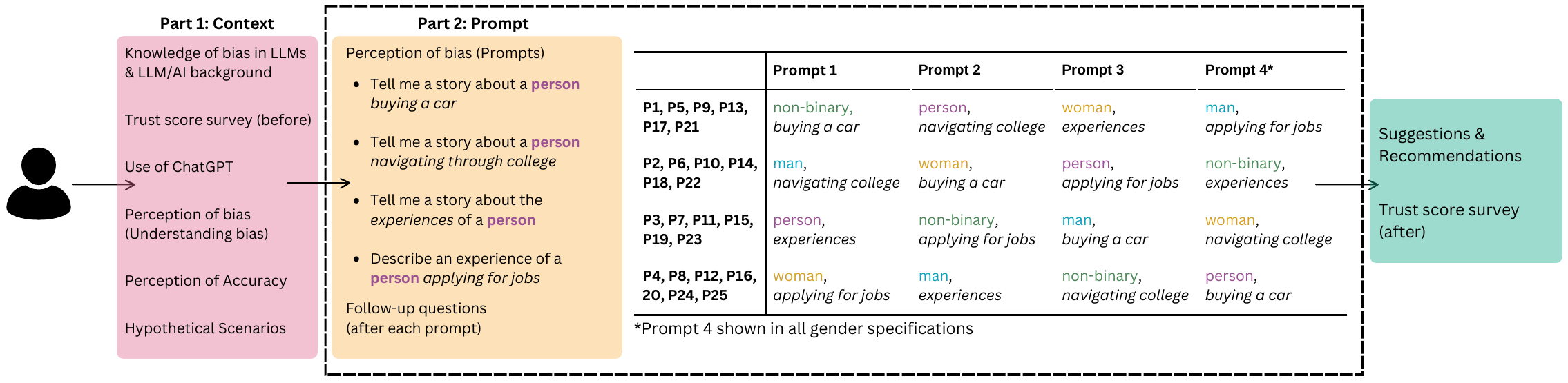}
    \caption{An overview of our study protocol}
    \label{fig:studyoverview}
\end{figure}

After showing each prompt, the interviewer asked the participants to share their immediate reactions and thoughts (in case they had not already). The interviewer then guided the discussion by asking if the participants thought ChatGPT was making assumptions about gender, if the statement could be perceived as offensive or harmful, why the participants thought ChatGPT responded the way it did, if they found the response surprising, and what follow-up questions they would ask ChatGPT if they had the option.
Each interview ended with a summary discussion of how the participant felt about ChatGPT or LLMs in general and for suggestions to improve its ability to discuss gender appropriately (Figure~\ref{fig:studyoverview}).

\subsection{Pilot Studies}
We conducted two Zoom-based pilot studies to refine the user study process and address potential issues. Based on participant feedback, we extended the study duration from 60 to 75 minutes, adjusted compensation, and decided to display all gender specifications for the fourth prompt to allow direct bias assessments. Additionally, we removed redundant questions, clarified the trust score survey, and resolved ambiguities in the study design.

\subsection{Conducting and analyzing interviews}
We conducted Zoom interviews with participants from February to September 2024, averaging 65 minutes each, and compensated them with a \$30 Amazon gift card. Interviews were transcribed using Zoom and analyzed through a collaborative coding process. In total, we collected and transcribed $\sim$$1,400+$ minutes of video and audio recordings using Zoom's transcription services. The details of our analysis are presented below.

We began our analysis by open coding all 25 transcripts following Corbin and Strauss’s grounded theory techniques~\cite{corbin2008basics}, developing an initial codebook to address each research question. Our approach was iterative and followed Braun and Clarke’s reflexive thematic analysis framework~\cite{braun2021thematic}, which includes familiarization with the data, coding, generating initial themes, reviewing and refining themes, and writing up. The first author led the coding process by reading all transcripts and drafting the initial codebook. This was followed by multiple collaborative meetings where all authors iteratively refined the codebook - shifting from descriptive codes to more interpretive, latent themes. Throughout the year-long data analysis period, the first author regularly discussed emerging themes and relationships with co-authors. Once we reached consensus on a final set of themes, the first and third authors reviewed all codes and themes for alignment with the data, ensuring conceptual clarity and eliminating redundancy.

We opted not to calculate the inter-rater reliability (IRR) in our analysis because it does not align with the interpretive framework foundational to qualitative research, as argued by McDonald et al. \cite{irrqualitative}. Our approach prioritized collaboration and consensus in the development of the codebook, where interpretations of the responses of the participants were collectively discussed and agreed on. Adding a numerical comparison of the coding choices after the fact would not have enhanced the rigor of our process. Instead, we focused on ensuring analytical rigor through in-depth discussions and careful refinement of our coding framework.



\subsection{Background: ChatGPT Usage} 
Participants reported using ChatGPT in various scenarios, mainly for writing tasks such as email writing (\peight{}, \peleven{}, \ptwentyfour{}), writing cover letters (\pnine), essays (\peleven), speeches (\pten{}), summaries (\pseven{}), and academic paper writing (\pone{}, \pfive{}). Although most found it effective, some noted repetitive output and the need for multiple prompts to refine results. For example, \pnine{} shared, \myquote{I have to keep telling it that it's wrong and then it apologizes, but keeps giving the same answer}.
Participants also used ChatGPT to create or refine resumes (\ptwo{}, \ptwenty{}), often finding it helpful. However, \pfifteen{} highlighted using it specifically to pass applicant tracking systems, \myquote{I use ChatGPT if I think a computer is going to look at it first}. For programming tasks (\ptwelve{}, \pseventeen{}, \ptwentythree), participants found it reliable for smaller problems but less effective for complex or niche tasks, as \ptwentythree{} explained, \myquote{It works for very small lines of code, but for very specific or niche queries, it doesn't}.
Brainstorming (\pthree{}), creating assignment outlines (\pthirteen{}, \pfourteen{}), and idea generation (\ptwentytwo{}) were also common uses, but participants reported limited success in these areas. \pthree{} noted redundancy in responses despite varied prompts, \myquote{I kept asking different questions, but I got the same answers}. Similarly, \ptwentytwo{} observed limitations in addressing recent topics, \myquote{I do not get references for newly published topics}.
Some participants used ChatGPT recreationally (\pseven{}) or for casual conversations (\psixteen{}, \ptwentyfive{}). For example, \pseven{} said, \myquote{I've used it for summarizing stuff and recreational purposes}. However, \ptwenty{}, who used ChatGPT to generate culturally relevant images, reported inaccuracies, \myquote{I've never received a single photo where the word is in Bengali}.
In general, the participants found ChatGPT most effective for structured and straightforward tasks but highlighted challenges with creativity, specificity, and handling recent or culturally nuanced topics

\subsection{RQ2: Perception of bias in LLMs}
\label{sec:rq2perceptionbias}

\subsubsection{Participants' Understanding \& Characterization of Bias}
We began by asking them to reflect on instances where they had encountered biased responses. We then asked how they discovered these biases, and whether they could identify biased responses. Participants were also asked about their specific expectations of \gpt and if their current experiences matched these expectations. We also asked if they found certain types of prompts to elicit more biased responses compared to others and if these biases affected their work or life in any way.
From the $25$ participants in the study, $2$ reported that they did not encounter bias during their interactions with LLMs, attributing this to their limited use of technology. The remaining $23$ participants shared a variety of experiences, from subtle biases to more noticeable patterns in \gpt behavior. Four primary themes emerged from their responses regarding perceived bias.

\pheading{Stereotypical bias:} 
The most commonly observed bias was stereotypical, particularly regarding gender assumptions. Participants shared several examples where \gpt displayed such biases. \pone{} noted that when asked to draw an arched nose, \gpt produced only Eurocentric features, \myquote{it could only draw the same like very eurocentric features... why can you not draw facial features beyond a very eurocentric ideal?}.
\pnine{} highlighted gender stereotypes in language, with \gpt associating terms like \say{beautiful dress} and \say{grace} with female CEOs, and \say{leadership} to male CEOs, \myquote{it was giving me things like beautiful dress instead of things you usually expect in a CEO like charge and making decisions}. Similarly, \pten{} noted that \gpt referred to male leaders as \say{Mr. President} but used \say{Miss XYZ} for women in similar roles, \myquote{for my mom, a president, it just wrote Miss Xyz}.

Bias against non-binary individuals was also observed. \pfifteen{} reported misgendering, with \gpt replacing their androgynous name with traditionally feminine ones, \myquote{it kept substituting random women's names for mine}. \ptwentyfive{} noted that \gpt adopted a passive-aggressive tone after recognizing their non-binary identity, \myquote{it started giving us different answers and got a tone... even refusing to answer some questions on gender}.

\pheading{Favoring of information:} The second theme that emerged involved the tendency of \gpt to favor certain types of information, such as widely accepted or historical knowledge, leading to responses that participants described as \say{prefers historic data} or \emph{stereotyping of information sources} as noted by \pnineteen{} who said \myquote{I would love to use the word stereotyping of information sources. It's not giving the diversity that is required for information because I believe every information put out there has a part}.

\pheading{Writing behavior:} A third theme centered on the \gpt's language processing or writing behaviour, where it often introduced unnecessary formal logic or mathematical explanations, and corrected colloquial or slang language without request. \peleven{} observed, \myquote{it'll kind of try to correct you if you use that (slang) in your writing}. 

\pheading{Indirect bias encounters:} 
Some participants, though not directly encountering bias, were aware of it through external sources such as Reddit or friends. For example, \pfive{} noted that prompts like generating an astronaut image often defaulted to American astronauts with U.S. flags or depicted prisoners as Black individuals. Additionally, they highlighted Google Translate’s gender assumptions when translating Bangla, where neutral pronouns were replaced with stereotypical roles, such as \myquote{he is playing football} and \myquote{she is cooking.} These observations reflect participants’ awareness of bias through experiences with other software.

When asked about recognizing biases in \gpt responses, most participants identified biases due to their familiarity with the topics, as \psixteen{} noted, \myquote{I usually ask about things I'm proficient with.} Some reported difficulty detecting subtle or situational biases, with \ptwentythree{} explaining that their lived experiences as an oppressed group made sexism and racism more noticeable. Others, like \ptwentytwo{}, remarked that bias is subjective, \myquote{unfair to me may be fair to you.} 
Most participants could distinguish hallucinations of \gpt from biases, although vague or gender/race-related prompts revealed stereotypical assumptions, as \pseventeen{} observed, \myquote{vague prompts makes the model rely on assumptions.}
When asked if they thought \gpt displayed bias toward certain types of prompts, the participants generally responded that bias was more apparent when the prompt was vague or related to gender or race. \pseventeen{} commented, \myquote{I think if the prompt is kind of vague or basic, I mean the model has to, I think, make assumptions, and then come up with something}.

When asked if they expected \gpt to be biased, responses varied. Some were surprised to discover bias, as \pthirteen{} noted, \myquote{it has to use existing data, which can be biased.} Others found it better than expected depending on the prompts, with \pnine{} sharing, \myquote{it depends on how you're using it.} Some expected bias from the start, such as \ptwentytwo{}, who noted, \myquote{real-life data is biased, and models trained on it will reflect that.} Overall, participants acknowledged bias as a persistent issue in AI systems.

\subsubsection{Prompt Findings}
To better understand perceived bias, we analyzed our findings from the prompts shown to the participants, organizing them based on the perspectives of men, women, and non-binary/transgender individuals. This analysis includes their immediate reactions, their perceptions of gender assumptions, and their views on harmful or offensive content in gendered (\textcolor{ManColor}{man}, \textcolor{WomanColor}{woman}, \textcolor{NonbinaryColor}{non-binary}) and neutral (\textcolor{PersonColor}{person}) prompts.

\input{Tables/promptfindings}

\pheading{Non-binary/Transgender Perspective:}

\noindent \textbf{\textcolor{NonbinaryColor}{Non-binary} Prompts:} 
Non-cisgender participants expressed frustration and, at times, outrage at ChatGPT's handling of non-binary representation, criticizing its responses as \emph{lacking nuance}, \emph{condescending}, \emph{cis-centric}, and \emph{stereotypical}. \psixteen{} shared, \myquote{It does sound condescending... it feels like, okay, I'm non-binary, but I'm as good as a man or a woman.} \pfifteen{} described job-related responses as \myquote{This is so cis-centric. I mean... how do you make a cis computer?... We're reducing the identity of a non-binary person to a struggle because of their gender}.
Participants also felt that ChatGPT assumed too much about gender, focusing on marginalization rather than individuality. \pone{} noted, \myquote{The word choice runs with the idea of a marginalized identity, focusing on that specifically}. Others found responses offensive or harmful, with \pfifteen{} stating, \myquote{I'm definitely offended. As someone from this community... it feels depersonalizing, turning the person into an object of struggle rather than acknowledging their humanity}. Some viewed the responses as contrived or damaging, as \pthree{} remarked, \myquote{It just feels like it could be perceived as an insignificant problem. People who aren't trans might even adopt the narrative that trans people are overly sensitive}. Meanwhile, \ptwo{} found the responses inauthentic or unintentionally amusing, \myquote{It creates problems in work and academic environments... not necessarily offensive, but it doesn't feel authentic at all}.

\noindent \textbf{\textcolor{WomanColor}{Woman} prompts:}
Non-binary participants criticized \gpt's responses as \emph{blatant}, \emph{uncertain}, \emph{emotional}, or \emph{AI-generated}. \ptwo{} remarked, \myquote{Discussing the...emotional side of things could be construed as a bit stereotypical}. \pfifteen{} noted uncertainty in the portrayal of the female character, \myquote{There's an uncertainty to her... It assumes that women are less powerful}. \pthree{} observed that \gpt assumed the female character was cisgender, \myquote{If Grace weren’t cis, there’d be a more complicated story}.
While most participants 
did not find the responses offensive, they shared concerns that associating women with traditional traits could reinforce stereotypical and less powerful images of women.

\noindent \textbf{\textcolor{ManColor}{Man} prompts:}
Non-binary participants noted that \gpt reinforced traditional gender roles and used \emph{typical male representation} with \emph{bigger words}. \psixteen{} commented, \myquote{If you remove pronouns and names, it’s clearly about a man... typical traits like \textit{engine roared}}. \pfifteen{} criticized the response as overly westernized and consumerist, \myquote{It’s so American... ``the drive towards the horizon''... ``a declaration of independence''}. \pfourteen{} noted the use of \emph{white-coded names} like Alex, while \pthree{} pointed towards the cis-centric nature of the model and remarked, \myquote{If you prompt, it just seems like they won't give any trans, right? It would never give a trans character unless explicitly asked for it}.
Participants commonly found the responses \emph{lacking diversity} or even \emph{humorous}, though not offensive. However, \pfourteen{} and \pfifteen{} expressed concerns that the statements could harmfully underemphasize women’s career ambitions or overemphasize masculine traits.

\noindent \textbf{\textcolor{PersonColor}{Person} prompts:}
Regarding the \textcolor{PersonColor}{person} prompts, non-binary participants were divided. Some appreciated the absence of gendered assumptions, while others felt that neutrality overlooked diverse gender experiences. Many described the responses as \emph{neutered} compared to gender-specific ones. \pone{} remarked, \myquote{It's not as gendered as the man or woman versions... just a standard approach.} \pone{} and \pfourteen{} noted the exclusion of non-binary and transgender identities, \myquote{It could have written this story about the non-binary person.} Three participants (\pthree{}, \pfour{}, \pfifteen{}) were offended by the lack of diversity. \pfifteen{} criticized the responses as generic and dismissive, \myquote{These robots are sexist... this is neutral... the others were much more specific}.


\pheading{Men's Perspective:}

\noindent \textbf{\textcolor{NonbinaryColor}{Non-binary} prompts:}
Men participants commonly found the responses biased, focusing heavily on the non-binary identity. \pseventeen{} remarked, \myquote{The emphasis is way more on the non-binary part than, you know, buying the car}, while \psix{} noted, \myquote{It associated challenges to non-binary people}. \ptwenty{} cautioned that they could be harmful depending on the context, \myquote{It might be harmful if someone younger reads it and thinks it has to be extremely tough}. However, two participants did not find anything problematic, stating that the story was \emph{realistic} or considered it to be \emph{a perfect story line}. 

\noindent \textbf{\textcolor{WomanColor}{Woman} prompts:}
Men thought that these responses were \emph{mellow}, \emph{creative}, or \emph{quite perfect} in some cases. \peighteen{} said, \myquote{I mean, you told her to write a story about a woman buying a car, so I don't think there were any assumptions which were wrong}. Three participants noted \gpt to show some bias or make assumptions, such as associating women with \emph{biology major} or \emph{color red/pink}.
None of the participants found the responses offensive or harmful. However, three participants (\pfive{}, \peight{}, \ptwenty{}) expressed \gpt was using a different tone like being assumptive or stereotypical.

\noindent \textbf{\textcolor{ManColor}{Man} prompts:}
Men participants observed \gpt's use of ``strong'' language, associating men with sporty cars and computer science majors. \pseven{} noted, \myquote{It goes in a more positive direction, avoiding struggles.} \pone{} highlighted how applying for jobs was framed as \myquote{a man going to war}. Some participants noted tonal differences across genders, but none found the responses offensive, though a few flagged concerns about inaccuracies, such as focusing on irrelevant details when buying a car.

\noindent \textbf{\textcolor{PersonColor}{Person} prompts:}
Men participants thought these responses were \emph{typical}, \emph{flowery}, or \emph{well constructed}. They also noticed that \gpt assigned a gender to the person's character in the story. \peighteen{} stated, \myquote{I don't think.. it has to choose a gender.. It doesn't matter if the model chooses a male as a subject or a female... If there is some unequal distribution, then there might be some sort of bias}.
They did not perceive the responses as offensive or harmful. However, participants felt that it made stereotypical gender assumptions depending on the gender it gave to the person character, maybe just not as strongly as the gender-specified prompts.

\pheading{Women's Perspective:}

\noindent \textbf{\textcolor{NonbinaryColor}{Non-binary} prompts:}
Women were critical of how non-binary individuals were portrayed, noting that \gpt often focused excessively on gender identity and reinforced stereotypes. \ptwentyone{} remarked, \myquote{It’s consumed just in that identity, not talking about their strength for the job}. Similarly, \ptwentythree{} criticized overly dramatic language, \myquote{They’re just non-binary, not a mythical creature... they’re just trying to take a subject}. They also noted inconsistency in pronoun usage for the Taylor character.
While majority of the participants didn’t find the responses offensive, some felt unqualified to judge as they didn’t identify as non-binary. Sharing the sentiments of other participants, \peleven{} commented that the identity-focused narrative could spread false information about non-binary people that everything is associated with their identity. 

\noindent \textbf{\textcolor{WomanColor}{Woman} prompts:}
Women reported to have felt uncomfortable with stereotypical portrayals, particularly when characters were shown as \emph{emotional}, \emph{hesitant}, or exhibiting \emph{traditional feminine traits}. \ptwentythree{} remarked, \myquote{The woman is very like small and shy, and they love to read, and they sit quietly... but women have different personalities, they have different characteristics}. Some noted how prompts often focused on themes like ``newfound independence'' with \ptwentyfour{} commenting, \myquote{For women, it’s independence; for non-binary people, it’s their identity.}
While participants were not offended, many expressed a need for portrayals that avoided stereotypes. \ptwentyfour{} added, \myquote{I’d want it to change its tone, incorporating gender without reinforcing stereotypical biases}.

\noindent \textbf{\textcolor{ManColor}{Man} prompts:}
Women participants commented that \gpt used \emph{big words} for men, such as \emph{adventurous} or \emph{stoic}. They reflected on how men were depicted as confident and in control. \pten{} also shared how there was \emph{never a moment of hesitation} in these responses. Participants also noticed that these responses were interesting, set in the current age, or went into greater depth.
The participants did not feel offended; however, they shared similar concerns as above about \gpt making strong assumptions by taking gender stereotypes into account.

\noindent \textbf{\textcolor{PersonColor}{Person} prompts:}
Women, like men and non-binary participants, found the responses to be \emph{generic} or \emph{vague}, often noting that \gpt assigned a gender to the ``person'' character and tailored responses accordingly. Many highlighted differences in how genders were portrayed, with \pten{} observing, \myquote{for women, the story moves toward independence, but men are assumed to start with it — they don’t have to fight for that part}. While participants did not find the responses offensive, they expressed concerns about the perpetuation of traditional gender narratives.

Participants reflected on potential reasons for these patterns, citing factors such as the limited availability of resources on non-binary identities online, biased portrayals in media, and the model's reliance on internet data that reflects societal biases. They also pointed to the lack of diverse data, including limited name diversity, which restricts the model's ability to represent gender-diverse identities accurately.

Overall, participants identified significant gender biases in \gpt's responses, with men depicted as strong and women as emotional or hesitant. Non-binary characters were often reduced to identity struggles, which many found reductive. Non-binary participants described responses as condescending and stereotypical, while women criticized outdated tropes, and men noted a lack of diversity but found responses less problematic.
Neutral prompts often reflected subtle gender stereotypes, as \gpt assigned a gender to ``person'' characters. 
We conclude that the perception of bias was subjective, shaped by the experiences of the participants and the familiarity with the topic. While responses were not seen as overtly offensive, participants highlighted the harm of reinforcing stereotypes, particularly for marginalized groups, and called for more diverse, nuanced training data to address these issues.

\subsection{RQ3: Perception of Accuracy in LLMs}
\label{sec:rq3perceptionaccuracy}

Participants were asked to recall instances when \gpt provided incorrect answers, reflect on its accuracy, and describe patterns in its performance. 
Participants across all gender groups shared similar concerns about the correctness and performance of \gpt, highlighting that errors often occurred with highly specific or technical topics. 
Most participants indicated that they could generally identify when \gpt's responses were accurate or not, especially if they were familiar with the subject, as \pthirteen{} noted, \myquote{If it's something I was already researching, I realize it's making an incorrect point}. However, for some participants, the verification of the correctness appeared more uncertain. \peleven{} said, \myquote{Sometimes I can tell when it's incorrect because I feel like that was a weird way to say it}. This suggests that familiarity with the topic plays an important role in the ability of a user to assess the accuracy of \gpt's responses.

When asked how they knew whether ChatGPT was correct, participants provided a range of strategies. Some relied on external verification, such as checking the information through search engines or comparing it to known sources. \pone{} explained, \myquote{I’ll do a little bit of research, like look it up on Google.}. Others noted that they had an \emph{intuition} (\pthirteen{}) or developed a sense of when responses felt logical or reasonable.
Participants observed that ChatGPT performed better on \emph{simple}, \emph{logical}, or \emph{generic} prompts but struggled with creative tasks. \pthirteen{} noted, \myquote{It seems more correct for logical or factual questions compared to subjective or societal analysis-type prompts} and \peighteen{} explained, \myquote{Mostly for programming things where there's some sort of creativity required... or some sort of design or decisions that need to be made... that's where ChatGPT usually gets it wrong}.
Expectations for accuracy varied. Some initially expected high accuracy but became cautious after noticing errors. \pthirteen{} said, \myquote{At first, I expected it to be accurate... but it can mislead you.} Others had modest expectations but were impressed by its improvement over time, with \ptwelve{} remarking, \myquote{Part of the reason why I didn't hop on the train at first is because… expectations were pretty low... With a well-engineered prompt, my confidence is pretty high nowadays}.

\subsection{RQ4: Perceived Trustworthiness of LLMs}
\label{sec:rq4perceptiontrust}

\subsubsection{Qualitative Analysis - Hypothetical Scenarios}
\label{sec:Trust_qual}
\hspace{0pt} \\
We asked participants whether they would use ChatGPT in hypothetical high-stakes scenarios involving financial and health outcomes, drawing inspiration from~\cite{humansaicontext} to introduce high stakes into LLM adoption decisions. These scenarios provided insights into how participants’ trust assessments and decision-making varied across different usage contexts. The scenarios and corresponding findings are detailed below.

\begin{itemize}
    \item \textbf{Health scenario:} Suppose you find yourself sick and go to visit a doctor. The doctor is not sure about the diagnosis. Would you recommend ChatGPT to identify the symptoms and help with the diagnosis so that the doctor can quickly determine the course of treatment?
\end{itemize}

Most participants (17 out of 25) were reluctant to trust ChatGPT for medical diagnoses, viewing it as suitable for general information but not critical decisions. As \pseven{} noted, \myquote{I don't think ChatGPT, or any language models should be used to for very critical purposes. For example, health care is something very critical which can actually affect life, so I don't think generally should be used, especially because their responses always seem very compelling, and that can actually leave doctors to.. trust them implicitly}. Another common theme was the lack of human interaction in \gpt's responses and it's inability to physically examine symptoms, with \ptwo{} stating, \myquote{I’d rather rely on a doctor who can observe and interact with me.} Participants emphasized that accurate diagnoses require physical examinations, which ChatGPT cannot perform. \pthree{} observed, \myquote{ChatGPT can’t see symptoms like rashes, which might change the diagnosis.} These findings underscore the limitations of text-based AI in addressing the complexities of healthcare. This sentiment highlights a broader concern about the limitations of text-based AI and the importance of human contact in healthcare decision-making.

Although majority of the participants were reluctant to use ChatGPT for medical diagnosis, some were open to using it for an initial evaluation of symptoms but not for complete diagnoses. \peighteen{} explained, \myquote{I don't really trust something like ChatGPT to help with diagnosis, maybe like a low level diagnosis, like if I have some headache or some cold. I can ask it why I got that}. A smaller group (2 participants) mentioned they might trust ChatGPT more if it were specifically trained on medical data. As \pnine{} noted, \myquote{I wouldn't recommend it as is, but if it's an LLM that is specifically trained for medical conditions, then maybe}.

\begin{itemize}
    \item \textbf{Financial scenario:} Suppose you are participating in a game show where you can win or lose money based on how well you respond to general knowledge questions. You can only use one resource among ChatGPT, books (e.g., field guides), search engines (e.g. google), online community), and so on. Which resource would you use? Does your answer change depending on certain factors?
\end{itemize}

A majority of participants (13 out of 25) preferred search engines, books, or online communities over ChatGPT, citing its unreliable output and the ease of verifying information through trusted sources. \ptwo{} emphasized, \myquote{I prefer verified, published information from reputable sources, like academic journals}. For these participants, search engines, books, and online communities were seen as more reliable and verifiable sources, which was crucial in a game show environment where mistakes could result in financial losses.

Conversely, 6 participants trusted ChatGPT for quick, and concise answers. \pfive{} highlighted its efficiency, \myquote{It's good at summarization... It would save me a lot of time that I would spend while searching on Google or any other search engine and it would do the same thing for me in 5 seconds}. Several participants trusted \gpt to answer general knowledge questions, particularly when the task was time-sensitive. For these participants, ChatGPT's ability to provide quick and concise responses made it an attractive option in a fast-paced, high-pressure situation. \psix{} mentioned, \myquote{If I had a very limited amount of time, maybe like 10 seconds or 20 seconds to answer, I'd go with ChatGPT}.

Overall, these findings suggest that while ChatGPT is appreciated for speed and convenience, its trustworthiness in high-stakes decisions remains questionable. Concerns about verification, its lack of real-world capabilities (e.g., physical exams in healthcare), and the need for domain-specific training limit its use in critical contexts. Overall, participants view ChatGPT as a supplementary tool, relying on human expertise and verifiable sources for dependable decision-making.

\begin{figure*}[!thb]
  \centering
    \includegraphics[width=\linewidth]{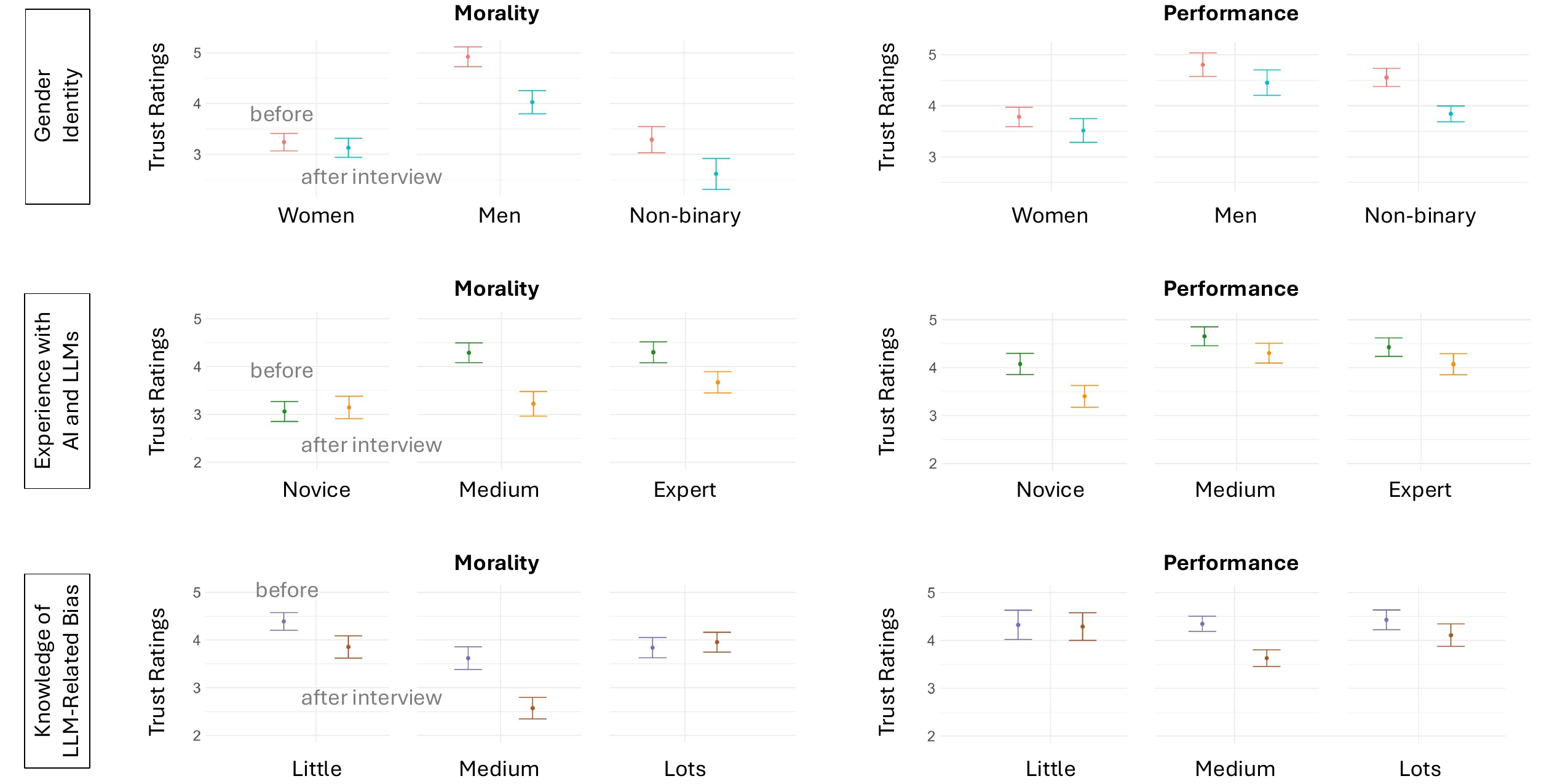}
    \caption{Before and after the interview trust ratings by self-reported gender, by the participant's expertise/bakcground in LLM and AI tools, and by the participants knowledge of bias in AI and LLMs}
    \label{fig:combine_genderbiasbackground}
\end{figure*}




\subsubsection{Quantitative Analysis - Trust Score} \hspace{0pt} \\
This section examines participants' self-reported trust in the LLM, categorized into two types: morality-based trust, reflecting perceptions of the LLM's benevolence, transparency, and ethicality, and performance-based trust, capturing perceptions of reliability and competence~\cite{malle2023measuring}. The findings offer a quantitative perspective on how participants perceive the trustworthiness of LLMs across these dimensions.

\paragraph{Gender and Trust}
We first examined how trust ratings varied by gender using a general linear model that included trust type (moral or performance-based), participants' gender (men, women, non-binary/transgender), and the time the trust ratings were collected (before and after the interviews).
Figure~\ref{fig:combine_genderbiasbackground} shows the results. 
We observed a significant effect of gender (${\chi}^2$ = 70.55, p < 0.001), with men reporting higher levels of trust in both the morality and performance of LLMs compared to women and non-binary participants. For performance-based trust, non-binary participants reported significantly higher levels of trust than women. This divergence in trust levels among non-binary participants between morality-based and performance-based trust contributed to an interaction effect between gender and trust type. Specifically, non-binary participants exhibited greater trust in LLM performance (significantly higher than women) but lower trust in LLM morality (comparable to women).
We also found a significant effect of the type of trust (${\chi}^2$ = 20.401, p < 0.001), such that participants reported higher levels of performance-based trust than morality-based trust.
Trust ratings decreased significantly after the interviews (${\chi}^2$ = 15.80, p < 0.001). However, we cannot make causal conclusions regarding whether this is driven by the deep reflection induced by the interview study or potential experimenter demand. 



\vspace{-7mm}
\paragraph{Experience with LLMs and Trust}
Following a similar approach, we used a general linear model to examine the effect of the participant's expertise in AI and LLMs on their trust in LLMs. Specifically, the linear model predicted trust ratings based on trust type, time (as in the previous model), and participants' self-reported expertise in AI and LLM tools, as described in section~\ref{sec:participantrecruitment}. 
The interactions between these factors were also included in the model. 
The results are presented in Figure~\ref{fig:combine_genderbiasbackground}.
As in the previous model, we found a significant effect of the type of trust and time (${\chi}^2$ = 17.69 and 15.81, p < 0.001). 
Furthermore, there was a significant effect of the experience and background of the participants in AI and LLM tools (${\chi}^2$ = 26.68, p < 0.001) such that individuals with higher self-reported expertise tended to trust LLMs more in both morality and performance. 
In contrast, novice AI users tended to trust LLMs less, possibly due to a lack of familiarity with how such systems operate or a higher degree of skepticism stemming from limited exposure to their capabilities and limitations. For example, \pthree{} noted that one of the reasons why they won't use \gpt in a high-stake health scenario (\ref{sec:Trust_qual}) was because of their limited experience with using LLMs.

\vspace{-5mm}
\paragraph{Knowledge of LLM Biases and Trust}
We constructed another general linear model to examine the effect of the knowledge of participants about LLM-related biases on their trust in LLMs. The linear model predicted trust ratings based on trust type, time (pre- and post-interviews), and participants' self-reported knowledge of biases in LLM outputs, as described in the methodology section. 
The interactions between these factors were also included in the model. The results are presented in Figure~\ref{fig:combine_genderbiasbackground}.
As in the previous model, we found a significant effect of the type of trust and time (${\chi}^2$ = 18.82 and 13.27, p < 0.001). 
Furthermore, there was a significant effect of the knowledge of participants about LLM-related biases (${\chi}^2$ = 21.82, p < 0.001), with individuals reporting greater knowledge of biases that tend to trust LLMs less in both morality and performance.
Interestingly, there was also a significant interaction between self-reported knowledge of LLM-related biases and the reporting time of trust rating (${\chi}^2$ = 7.76, p < 0.001). 
Participants with minimal or extensive knowledge of LLM biases showed negligible changes in their trust ratings for both morality and performance over time. 
However, the most notable group was participants who self-reported a medium level of knowledge about biases. Their trust in LLMs, both in morality and performance, decreased significantly after the interview. These findings corroborate existing work where participants' knowledge of their domain in a computer vision AI application had a huge influence on their trust~\cite{humansaicontext}.
We hypothesize that the interview questions, which prompted participants to critically reflect on the behaviors and outputs of LLMs, had a pronounced effect on this group. 
These participants likely had enough working knowledge to recognize the concept of bias, but lacked the depth of understanding necessary to contextualize how it manifests in LLMs, leading to a change in their trust.\newline


\vspace{-2mm}
\noindent \textbf{Key takeaways:} 
Our findings highlight significant effects of gender, AI expertise, and LLM experience on trust. Men reported higher trust overall, while non-binary participants showed greater trust in LLM performance but less in morality, similar to women. Higher AI expertise was linked to greater trust, whereas novice users expressed lower trust, likely due to unfamiliarity or skepticism. Knowledge of LLM biases generally reduced trust, with participants possessing medium knowledge experiencing the largest decline after reflecting critically during interviews. 
These results emphasize how deep reflection and awareness of biases influence trust perceptions across diverse groups.

\section{Discussion}
In this section, we first present the suggestions and recommendations made by participants for improving how gender can be represented in LLMs more appropriately. We then outline a set of design implications informed by these insights, highlighting opportunities for sociotechnical systems like LLMs that better reflect and respect diverse social experience.

\subsection{RQ5: Recommendations from Participants}
\label{sec:rq5recommendations}

Participants shared a range of suggestions to enhance the ability of LLMs to address gender-related topics appropriately. 
A common theme was \textbf{the need for greater diversity in training data}. Participants emphasized the importance of incorporating real-life testimonies and lived experiences of individuals of various genders, cultures, and contexts. They argued that this would improve the model's understanding of gender nuances and promote a more equitable representation of voices. For example, \ptwo{} commented that \myquote{going out of the way to collect testimonies, experiences from like gender minorities would give it some benefits that it has a repository of real sources as opposed to just evaluating what little exists online}.

Participants also emphasized \textbf{the importance of responding to men, women, and non-binary individuals with equal depth and consideration} to ensure fairness while respecting individual identities. The participants noted that achieving a balance between tailored responses and generalization is crucial. This approach would ensure that, while responses respect individual identities, they also maintain consistency and fairness across diverse user interactions. For example, \pseven{} suggested, \myquote{make the machine understand what biases are and we have penalties inside this to account for that... what it should ensure is that, regardless of of what the gender or ethnicity background is - the outcomes stay uniform}. 

Several participants highlighted the potential for \textbf{LLMs to ask clarifying questions or seek additional context when engaging with users}. This proactive approach could help personalize interactions (\peighteen{}) and minimize the risk of misinterpretation, particularly when discussing sensitive topics such as gender. \pfive{} also pressed upon the ability of LLMs to give multiple options to choose from especially for sensitive contexts, \myquote{Anything that has to do with gender or has some ethical implications - it should somehow give multiple answers to choose from, or at least give multiple perspectives}. Participants also suggested that \textbf{responses could benefit from greater depth}, with some noting that overly surface-level or excessively positive responses often fail to capture the complexities of human experiences. They recommended that the system sound more natural and avoid weaving all narratives into an artificially positive outlook, which can feel inauthentic or dismissive of genuine challenges.

Transparency also emerged as a critical concern. Participants expressed the need for \textbf{LLMs to clearly communicate how they generate responses} and the limitations of their knowledge. \pfourteen{} shared, \myquote{maybe there needs to be more transparency... where it's pulling its sources from or... being more transparent about how, why, it's showing what it's showing}. By providing more insight into the data sources, algorithms, and design decisions that underpin these systems, LLMs could build greater trust and accountability among users. 
Additional recommendations included improving the model's handling of pronoun diversity and incorporating explicit safeguards through mechanisms such as a scoring function for guardrails.

\subsection{Future Opportunities \& Design Implications}
\subsubsection{Collaborative Gender-Inclusive Co-Design.}
Participants in our study suggested that including real-world testimonies and diverse life stories could improve representation. To address this, we propose a \textit{collaborative gender-inclusive co-design approach}~\cite{harringtonpd-cscw19, strangersatgate-dantec-cscw15}, where gender-diverse users are not merely testers but active contributors throughout the model development lifecycle. This participatory involvement means integrating their narratives, lived experiences, and feedback from the very beginning—during data curation, prompt engineering, and iterative evaluation. Co-design sessions with gender-diverse users may be able to surface nuanced biases early on in the model development lifecycle. Not only can this collaborative process empower marginalized voices but also ensure that LLMs reflect diverse gender perspectives with authenticity and respect.

\subsubsection{Contextual Clarification \& Transparent Explanaitions.}
A key concern raised by participants was the tendency of LLMs to make broad generalizations based on minimal input and with the opacity of LLM decision-making. To address these concerns, we propose integrating contextual clarification and transparent explanations into LLM interactions. This mechanism would enable the LLM to proactively seek additional user input when it detects vague or ambiguous prompts. For example, if a user asks, \say{Tell me about the engineer?} the system could respond with, \say{Would you like a gender-neutral perspective, or should I include gender-specific contexts?} or additional questions like \say{Which gender would you like me to assume and what pronouns should I use?}.
Additionally, the LLM could provide transparent or multiple explanations~\cite{duan-mitigatinggenderbias-xai} about how it constructed its responses, including the datasets referenced, and any other modic logic or confidence levels. These methods may allow users to trace the origin of information, understand potential biases, and critically evaluate the outout which may foster trust in the system.

\subsubsection{Gender Sensitive Prompt Test Suite for Bias Detection.}
Our study demonstrates how LLMs can exhibit biases in gendered responses, underscoring the need for systematic evaluation. In this work, we curated prompts that can potentially serve as a valuable test suite to evaluate how well an LLM handles gender in its responses. For example, a developer of a new LLM could use these prompts to assess whether the model lets gender influence its output or if it addresses inclusivity in an appropriate and neutral manner. This approach helps identify potentially harmful or inappropriate content early in the development process, ensuring that the LLM handles gender-related aspects effectively.

\subsubsection{Real-Time Bias Auditing.}
Our findings reveal that perceptions of bias in LLM responses vary significantly across gender identities. While some participants flagged certain responses as stereotypical or offensive, others viewed the same outputs as neutral or appropriate. This variation highlights a key challenge: bias in LLMs is not always universally recognizable, making it difficult to detect and address through one-size-fits-all evaluation methods. To address this, we propose a \textit{real-time bias auditing mechanism} where users can flag, annotate, and suggest corrections for biased or stereotypical responses during live interactions. These annotations can be collected and analyzed to identify patterns of bias, allowing developers to prioritize critical areas for model refinement. Additionally, community-driven auditing sessions can surface harmful biases that may not be evident during isolated testing, enhancing accountability and inclusivity through collective oversight~\cite{duboiscommunityqa}.

\section{Limitations and Future Work}
The study has several limitations that suggests directions for future research. 
First, it focused on a specific set of prompts framed in particular ways, which may not reflect the full spectrum of LLM responses. 
The study did not examine how models handle other gender identities or how varying prompt phrasing, including positive or negative tones, might influence outputs. Additionally, the non-interactive prompt presentation limited participants’ ability to engage directly with the model and provide feedback based on live interactions. Future studies could incorporate interactive sessions, enabling participants to explore model responses in real time and offer more nuanced feedback.

We also exclusively focused on ChatGPT, which may not represent the range of responses from other LLMs. Future work can compare multiple models to reveal broader patterns in how different architectures address gender bias, accuracy, and trustworthiness. 
Investigating how variations in training data and design choices influence inclusivity would also be valuable. Expanding the study to additional use cases beyond healthcare and general knowledge scenarios would help assess LLM performance in diverse contexts with varying stakes.

\section{Conclusion}
We explore how gender-diverse populations perceive the utility of LLMs using \gpt as a case study. 
We found that gender-specific prompts elicited identity-specific responses, with non-binary/transgender participants noting condescending and stereotypical portrayals. 
This study provides valuable insights on how gender-diverse populations perceive the utility of LLMs such as \gpt, particularly in relation to bias, accuracy, and trust.
Our findings suggest that gendered prompts evoke more identity-specific responses, with non-binary/transgender participants highlighting concerns around condescending and stereotypical portrayals.
While perceived accuracy remained relatively consistent across gender groups, overall participants noted that technical and creative tasks often led to errors. 
Trust in LLMs varied by gender, with men showing higher trust than women and non-binary participants, with non-binary participants trusting LLM performance more than morality. 
The findings highlight the need for diverse training data, equitable responses across genders, and greater transparency to foster inclusivity and trust in future LLMs.
\bibliographystyle{ACM-Reference-Format}
\bibliography{references}

%









\end{document}

%% file: commands.tex
\newcommand{\pheading}[1]{\smallskip\noindent\textbf{#1}}

\definecolor{ManColor}{HTML}{08A4BD}
\definecolor{WomanColor}{HTML}{D5A021}
\definecolor{NonbinaryColor}{HTML}{498756}
\definecolor{PersonColor}{HTML}{9B5094}

\newcommand{\pone}{\textcolor{NonbinaryColor}{P1}}
\newcommand{\ptwo}{\textcolor{NonbinaryColor}{P2}}
\newcommand{\pthree}{\textcolor{NonbinaryColor}{P3}}
\newcommand{\pfour}{\textcolor{NonbinaryColor}{P4}}

\newcommand{\pfive}{\textcolor{ManColor}{P5}}
\newcommand{\psix}{\textcolor{ManColor}{P6}}
\newcommand{\pseven}{\textcolor{ManColor}{P7}}
\newcommand{\peight}{\textcolor{ManColor}{P8}}

\newcommand{\pnine}{\textcolor{WomanColor}{P9}}
\newcommand{\pten}{\textcolor{WomanColor}{P10}}
\newcommand{\peleven}{\textcolor{WomanColor}{P11}}
\newcommand{\ptwelve}{\textcolor{WomanColor}{P12}}

\newcommand{\pthirteen}{\textcolor{NonbinaryColor}{P13}}
\newcommand{\pfourteen}{\textcolor{NonbinaryColor}{P14}}
\newcommand{\pfifteen}{\textcolor{NonbinaryColor}{P15}}
\newcommand{\psixteen}{\textcolor{NonbinaryColor}{P16}}

\newcommand{\pseventeen}{\textcolor{ManColor}{P17}}
\newcommand{\peighteen}{\textcolor{ManColor}{P18}}
\newcommand{\pnineteen}{\textcolor{ManColor}{P19}}
\newcommand{\ptwenty}{\textcolor{ManColor}{P20}}

\newcommand{\ptwentyone}{\textcolor{WomanColor}{P21}}
\newcommand{\ptwentytwo}{\textcolor{WomanColor}{P22}}
\newcommand{\ptwentythree}{\textcolor{WomanColor}{P23}}
\newcommand{\ptwentyfour}{\textcolor{WomanColor}{P24}}

\newcommand{\ptwentyfive}{\textcolor{NonbinaryColor}{P25}}

\definecolor{lightpink}{RGB}{237,157,202}
\definecolor{lightred}{RGB}{210,121,121}
\definecolor{lightorange}{RGB}{230,170,50}
\definecolor{lightgold}{RGB}{210,194,121}
\definecolor{lightgreen}{RGB}{121,210,121}
\definecolor{lightaqua}{RGB}{121,206,210}
\definecolor{lightblue}{RGB}{121,124,210}
\definecolor{lightpurple}{RGB}{153,102,255}
\definecolor{red}{RGB}{178,34,34}
\definecolor{gray}{RGB}{166,166,166}

\newcommand{\todo}[1]{\textbf{\textcolor{red}{TODO:}}}

\newcommand{\myquote}[1]{\emph{``#1''}}                         

\newcommand{\eat}[1]{\relax}
\newcommand{\gpt}{ChatGPT\xspace}

%% file: Tables/researchquestions.tex
\begin{figure}[t]
\caption{Research questions to understand people's perception of bias in a real-world application of large language models.}
\rowcolors{1}{gray!25}{white}
\small 
{\renewcommand{\arraystretch}{1.2} 
\setlength{\tabcolsep}{6pt} 
\begin{tabular}{p{0.29\columnwidth}p{0.66\columnwidth}} 
\toprule
\textbf{Research Question} & \textbf{Finding Summary} \\ 
\midrule
RQ1: How do gendered and neutral prompts influence ChatGPT's responses? & 
Our results indicate that gendered prompts elicit richer, more identity-specific responses from \gpt compared to neutral prompts, which are more generic. (Section~\ref{sec:rq1gptresponsepatterns}) \\
RQ2: How does the perception of bias in LLMs vary across gender diverse populations? & 
Our results indicate that perceived bias in LLMs varied, with non-binary/transgender participants noting condescending and stereotypical responses, women highlighting emotional and traditional portrayals, and men observing a lack of diversity but fewer concerns. (Section~\ref{sec:rq2perceptionbias})\\
RQ3: How does the perception of accuracy in LLMs vary across gender diverse populations? & 
Our results indicate that perceived accuracy in LLMs was similar across gender-diverse groups, with participants noting errors in technical topics and creative tasks. Most could assess accuracy when familiar with the subject, while others relied on external checks. (Section~\ref{sec:rq3perceptionaccuracy})\\
RQ4: How does the perceived trustworthiness of LLMs vary across gender diverse populations? &
Our results indicate that perceived trustworthiness of LLMs varied by gender, with men reporting higher trust, especially in performance. Non-binary participants showed higher performance-based trust but similar morality-based trust. (Section~\ref{sec:rq4perceptiontrust})\\
RQ5: What suggestions do users have for improving the way LLMs handle gender-related content? & 
Our results indicate that participants suggested improving LLMs by diversifying training data to include real-life gender experiences, ensuring equal depth in responses to all genders, incorporating clarifying questions, and the need for transparency in response generation. (Section~\ref{sec:rq5recommendations})\\
\bottomrule
\end{tabular}
}
\label{fig:rqs}
\vspace{-4mm}
\end{figure}




%% file: Tables/participantknowledge.tex
\begin{table}[!ht]
    \centering
    \caption{Participants' knowledge of bias in LLMs/AI and LLMs/AI background)}
    \label{tab:participantknowledge}
    \begin{tabular}{p{0.25\textwidth} p{0.17\textwidth} p{0.18\textwidth} p{0.2\textwidth}}
        \toprule
         & \textbf{Low-LLM/AI} & \textbf{Medium-LLM/AI} & \textbf{High-LLM/AI} \\ 
        \midrule
        \textbf{Low-bias in LLM} 
         & \textcolor{NonbinaryColor}{P4}, \textcolor{WomanColor}{P11} 
         & \textcolor{ManColor}{P19} 
         & \textcolor{ManColor}{P6} \\ 
        \textbf{Medium-bias in LLM} 
         & \textcolor{NonbinaryColor}{P2}, \textcolor{NonbinaryColor}{P3}, \textcolor{NonbinaryColor}{P13}, \textcolor{NonbinaryColor}{P14} 
         & \textcolor{ManColor}{P5}, \textcolor{NonbinaryColor}{P15}, \textcolor{WomanColor}{P24} 
         & \textcolor{ManColor}{P7}, \textcolor{WomanColor}{P10}, \textcolor{ManColor}{P17}, \textcolor{ManColor}{P18}, \textcolor{WomanColor}{P23} \\ 
        \textbf{High-bias in LLM}
         & \textcolor{WomanColor}{P12}, \textcolor{WomanColor}{P21}, \textcolor{NonbinaryColor}{P25} 
         & \textcolor{NonbinaryColor}{P1}, \textcolor{NonbinaryColor}{P16}, \textcolor{WomanColor}{P22} 
         & \textcolor{ManColor}{P8}, \textcolor{WomanColor}{P9}, \textcolor{ManColor}{P20} \\ 
        \bottomrule
    \end{tabular}
\end{table}

    

%% file: Tables/promptfindings.tex
\begin{table}[!ht]
    \centering
    \caption{Summary of Key Findings by Participant Perspective (Rows) and Prompt Type (Column)}
    \label{tab:perspectivefindings}
    \begin{tabular}{p{0.18\textwidth} p{0.18\textwidth} p{0.18\textwidth} p{0.18\textwidth} p{0.18\textwidth}}
        \toprule
        & \textbf{Non-binary} & \textbf{Woman} & \textbf{Man} & \textbf{Person} \\
        \midrule
        \textbf{Non-binary / Transgender Perspective} & Frustration with cis-centric responses, oversimplified struggles with gender identity. & Stereotypical emotional portrayal and assumptions about power. & Reinforcement of traditional gender roles and masculine traits. & Mixed views: Some liked neutrality, others felt it overlooked non-binary complexities. \\
        \midrule
        \textbf{Men's Perspective} & Focus on non-binary identity struggles, missing broader context. & Neutral, sometimes stereotypical portrayal, e.g., biology and color preferences. & Positive but traditional masculine traits lacking diversity. & Neutral prompts too generic, gender assumptions still present. \\
        \midrule
        \textbf{Women's Perspective} & Non-binary prompts reduce identity to struggles, lack of nuance. & Stereotypical emotional portrayals and power limitations. & Traditional masculine traits, not offensive, but lacking diversity. & Neutral prompts feel vague and unnuanced, lacked diversity. \\
        \bottomrule
    \end{tabular}
\end{table}

%% file: main.bbl

\begin{thebibliography}{72}


\ifx \showCODEN    \undefined \def \showCODEN     #1{\unskip}     \fi
\ifx \showISBNx    \undefined \def \showISBNx     #1{\unskip}     \fi
\ifx \showISBNxiii \undefined \def \showISBNxiii  #1{\unskip}     \fi
\ifx \showISSN     \undefined \def \showISSN      #1{\unskip}     \fi
\ifx \showLCCN     \undefined \def \showLCCN      #1{\unskip}     \fi
\ifx \shownote     \undefined \def \shownote      #1{#1}          \fi
\ifx \showarticletitle \undefined \def \showarticletitle #1{#1}   \fi
\ifx \showURL      \undefined \def \showURL       {\relax}        \fi
\providecommand\bibfield[2]{#2}
\providecommand\bibinfo[2]{#2}
\providecommand\natexlab[1]{#1}
\providecommand\showeprint[2][]{arXiv:#2}

\bibitem[SM(2025)]%
        {SM}
 \bibinfo{year}{2025}\natexlab{}.
\newblock \bibinfo{title}{Supplementary materials for LLM Bias}.
\newblock \bibinfo{howpublished}{\url{https://osf.io/ywmtg/?view_only=7a38b88af78e45ad85a33b70f88f717f}}.
\newblock


\bibitem[Abid et~al\mbox{.}(2021)]%
        {antimuslimbias}
\bibfield{author}{\bibinfo{person}{Abubakar Abid}, \bibinfo{person}{Maheen Farooqi}, {and} \bibinfo{person}{James Zou}.} \bibinfo{year}{2021}\natexlab{}.
\newblock \showarticletitle{Persistent Anti-Muslim Bias in Large Language Models}. In \bibinfo{booktitle}{\emph{Proceedings of the 2021 AAAI/ACM Conference on AI, Ethics, and Society}} (Virtual Event, USA) \emph{(\bibinfo{series}{AIES '21})}. \bibinfo{publisher}{Association for Computing Machinery}, \bibinfo{address}{New York, NY, USA}.
\newblock
\showISBNx{9781450384735}
\href{https://doi.org/10.1145/3461702.3462624}{doi:\nolinkurl{10.1145/3461702.3462624}}


\bibitem[Angwin et~al\mbox{.}(2016)]%
        {Angwin16}
\bibfield{author}{\bibinfo{person}{Julia Angwin}, \bibinfo{person}{Jeff Larson}, \bibinfo{person}{Surya Mattu}, {and} \bibinfo{person}{Lauren Kirchner}.} \bibinfo{year}{2016}\natexlab{}.
\newblock \showarticletitle{Machine Bias}.
\newblock \bibinfo{journal}{\emph{ProPublica}}  \bibinfo{volume}{May 23} (\bibinfo{year}{2016}).
\newblock
\newblock
\shownote{\url{https://www.propublica.org/article/machine-bias-risk-assessments-in-criminal-sentencing}}.


\bibitem[Ashwin et~al\mbox{.}(2023)]%
        {ashwin2023using}
\bibfield{author}{\bibinfo{person}{Julian Ashwin}, \bibinfo{person}{Aditya Chhabra}, {and} \bibinfo{person}{Vijayendra Rao}.} \bibinfo{year}{2023}\natexlab{}.
\newblock \showarticletitle{Using large language models for qualitative analysis can introduce serious bias}.
\newblock \bibinfo{journal}{\emph{arXiv preprint arXiv:2309.17147}} (\bibinfo{year}{2023}).
\newblock


\bibitem[Bartl and Leavy(2024)]%
        {leavy-bartl2024showgirls}
\bibfield{author}{\bibinfo{person}{Marion Bartl} {and} \bibinfo{person}{Susan Leavy}.} \bibinfo{year}{2024}\natexlab{}.
\newblock \showarticletitle{From'Showgirls' to'Performers': Fine-tuning with Gender-inclusive Language for Bias Reduction in LLMs}.
\newblock \bibinfo{journal}{\emph{arXiv preprint arXiv:2407.04434}} (\bibinfo{year}{2024}).
\newblock


\bibitem[Bender et~al\mbox{.}(2021)]%
        {dangersllms2021}
\bibfield{author}{\bibinfo{person}{Emily~M. Bender}, \bibinfo{person}{Timnit Gebru}, \bibinfo{person}{Angelina McMillan-Major}, {and} \bibinfo{person}{Shmargaret Shmitchell}.} \bibinfo{year}{2021}\natexlab{}.
\newblock \showarticletitle{On the Dangers of Stochastic Parrots: Can Language Models Be Too Big?}. In \bibinfo{booktitle}{\emph{Proceedings of the 2021 ACM Conference on Fairness, Accountability, and Transparency (FAccT)}}. \bibinfo{publisher}{ACM}, \bibinfo{pages}{610--623}.
\newblock
\href{https://doi.org/10.1145/3442188.3445922}{doi:\nolinkurl{10.1145/3442188.3445922}}


\bibitem[Binns and Veale(2022)]%
        {gender_nlg}
\bibfield{author}{\bibinfo{person}{Reuben Binns} {and} \bibinfo{person}{Michael Veale}.} \bibinfo{year}{2022}\natexlab{}.
\newblock \showarticletitle{Adhering, Steering, and Queering: Treatment of Gender in Natural Language Generation}. In \bibinfo{booktitle}{\emph{Proceedings of the Conference on Fairness, Accountability, and Transparency (FAT* 2022)}}. ACM, \bibinfo{pages}{142--155}.
\newblock


\bibitem[Braun and Clarke(2021)]%
        {braun2021thematic}
\bibfield{author}{\bibinfo{person}{Virginia Braun} {and} \bibinfo{person}{Victoria Clarke}.} \bibinfo{year}{2021}\natexlab{}.
\newblock \bibinfo{booktitle}{\emph{Thematic Analysis: A Practical Guide}}.
\newblock \bibinfo{publisher}{SAGE Publications Ltd}, \bibinfo{address}{London}.
\newblock
\showISBNx{978-1473953246}


\bibitem[Buolamwini and Gebru(2018)]%
        {buolamwini2018gender}
\bibfield{author}{\bibinfo{person}{Joy Buolamwini} {and} \bibinfo{person}{Timnit Gebru}.} \bibinfo{year}{2018}\natexlab{}.
\newblock \showarticletitle{Gender shades: Intersectional accuracy disparities in commercial gender classification}. In \bibinfo{booktitle}{\emph{Conf. on fairness, accountability and transparency}}. PMLR, \bibinfo{pages}{77--91}.
\newblock


\bibitem[Corbin and Strauss(2008)]%
        {corbin2008basics}
\bibfield{author}{\bibinfo{person}{Juliet Corbin} {and} \bibinfo{person}{Anselm Strauss}.} \bibinfo{year}{2008}\natexlab{}.
\newblock \bibinfo{booktitle}{\emph{Basics of Qualitative Research: Techniques and Procedures for Developing Grounded Theory} (\bibinfo{edition}{3rd} ed.)}.
\newblock \bibinfo{publisher}{SAGE Publications Ltd}, \bibinfo{address}{Thousand Oaks, CA}.
\newblock
\showISBNx{978-1412906449}


\bibitem[Crawford and Joler(2021)]%
        {gender_bias_nlp}
\bibfield{author}{\bibinfo{person}{Kate Crawford} {and} \bibinfo{person}{Vladan Joler}.} \bibinfo{year}{2021}\natexlab{}.
\newblock \showarticletitle{Harms of Gender Exclusivity and Challenges in Non-Binary Representation in Language Technologies}.
\newblock \bibinfo{journal}{\emph{AI \& Society}} \bibinfo{volume}{36}, \bibinfo{number}{1} (\bibinfo{year}{2021}), \bibinfo{pages}{245--261}.
\newblock


\bibitem[Dengel et~al\mbox{.}(2023)]%
        {dengel2023qualitative}
\bibfield{author}{\bibinfo{person}{Andreas Dengel}, \bibinfo{person}{Rupert Gehrlein}, \bibinfo{person}{David Fernes}, \bibinfo{person}{Sebastian G{\"o}rlich}, \bibinfo{person}{Jonas Maurer}, \bibinfo{person}{Hai~Hoang Pham}, \bibinfo{person}{Gabriel Gro{\ss}mann}, {and} \bibinfo{person}{Niklas Dietrich~genannt Eisermann}.} \bibinfo{year}{2023}\natexlab{}.
\newblock \showarticletitle{Qualitative research methods for large language models: Conducting semi-structured interviews with ChatGPT and BARD on Computer Science Education}. In \bibinfo{booktitle}{\emph{Informatics}}, Vol.~\bibinfo{volume}{10}. MDPI, \bibinfo{pages}{78}.
\newblock


\bibitem[Dietvorst and Bartels(2022)]%
        {trust_in_ai}
\bibfield{author}{\bibinfo{person}{Berkeley~J Dietvorst} {and} \bibinfo{person}{Daniel~M Bartels}.} \bibinfo{year}{2022}\natexlab{}.
\newblock \showarticletitle{The Value of Measuring Trust in AI - A Socio-Technical System Perspective}.
\newblock \bibinfo{journal}{\emph{AI \& Society}} \bibinfo{volume}{37}, \bibinfo{number}{2} (\bibinfo{year}{2022}), \bibinfo{pages}{111--126}.
\newblock


\bibitem[Dixon et~al\mbox{.}(2018)]%
        {biastextclassification}
\bibfield{author}{\bibinfo{person}{Lucas Dixon}, \bibinfo{person}{John Li}, \bibinfo{person}{Jeffrey Sorensen}, \bibinfo{person}{Nithum Thain}, {and} \bibinfo{person}{Lucy Vasserman}.} \bibinfo{year}{2018}\natexlab{}.
\newblock \showarticletitle{Measuring and Mitigating Unintended Bias in Text Classification}. In \bibinfo{booktitle}{\emph{Proceedings of the 2018 AAAI/ACM Conference on AI, Ethics, and Society}} (New Orleans, LA, USA) \emph{(\bibinfo{series}{AIES '18})}. \bibinfo{publisher}{Association for Computing Machinery}, \bibinfo{address}{New York, NY, USA}, \bibinfo{pages}{67–73}.
\newblock
\showISBNx{9781450360128}
\href{https://doi.org/10.1145/3278721.3278729}{doi:\nolinkurl{10.1145/3278721.3278729}}


\bibitem[Dong et~al\mbox{.}(2025)]%
        {dong-responsibleds-cscw25}
\bibfield{author}{\bibinfo{person}{Ziwei Dong}, \bibinfo{person}{Ameya Patil}, \bibinfo{person}{Yuichi Shoda}, \bibinfo{person}{Leilani Battle}, {and} \bibinfo{person}{Emily Wall}.} \bibinfo{year}{2025}\natexlab{}.
\newblock \showarticletitle{Behavior Matters: An Alternative Perspective on Promoting Responsible Data Science}.
\newblock \bibinfo{journal}{\emph{Proc. ACM Hum.-Comput. Interact.}} \bibinfo{volume}{9}, \bibinfo{number}{2}, Article \bibinfo{articleno}{CSCW034} (\bibinfo{date}{May} \bibinfo{year}{2025}), \bibinfo{numpages}{23}~pages.
\newblock
\href{https://doi.org/10.1145/3710932}{doi:\nolinkurl{10.1145/3710932}}


\bibitem[Draxler et~al\mbox{.}(2023)]%
        {draxler2023gender}
\bibfield{author}{\bibinfo{person}{Fiona Draxler}, \bibinfo{person}{Daniel Buschek}, \bibinfo{person}{Mikke Tavast}, \bibinfo{person}{Perttu H{\"a}m{\"a}l{\"a}inen}, \bibinfo{person}{Albrecht Schmidt}, \bibinfo{person}{Juhi Kulshrestha}, {and} \bibinfo{person}{Robin Welsch}.} \bibinfo{year}{2023}\natexlab{}.
\newblock \showarticletitle{Gender, age, and technology education influence the adoption and appropriation of LLMs}.
\newblock \bibinfo{journal}{\emph{arXiv preprint arXiv:2310.06556}} (\bibinfo{year}{2023}).
\newblock


\bibitem[Duan et~al\mbox{.}(2025)]%
        {genderstereotypesinai-25}
\bibfield{author}{\bibinfo{person}{Wen Duan}, \bibinfo{person}{Lingyuan Li}, \bibinfo{person}{Guo Freeman}, {and} \bibinfo{person}{Nathan McNeese}.} \bibinfo{year}{2025}\natexlab{}.
\newblock \showarticletitle{A Scoping Review of Gender Stereotypes in Artificial Intelligence}. In \bibinfo{booktitle}{\emph{Proceedings of the 2025 CHI Conference on Human Factors in Computing Systems}} \emph{(\bibinfo{series}{CHI '25})}. \bibinfo{publisher}{Association for Computing Machinery}, \bibinfo{address}{New York, NY, USA}, Article \bibinfo{articleno}{995}, \bibinfo{numpages}{20}~pages.
\newblock
\showISBNx{9798400713941}
\href{https://doi.org/10.1145/3706598.3713093}{doi:\nolinkurl{10.1145/3706598.3713093}}


\bibitem[Duan et~al\mbox{.}(2024)]%
        {duan-mitigatinggenderbias-xai}
\bibfield{author}{\bibinfo{person}{Wen Duan}, \bibinfo{person}{Nathan McNeese}, \bibinfo{person}{Guo Freeman}, {and} \bibinfo{person}{Lingyuan Li}.} \bibinfo{year}{2024}\natexlab{}.
\newblock \showarticletitle{Mitigating Gender Stereotypes Toward AI Agents Through an eXplainable AI (XAI) Approach}.
\newblock \bibinfo{journal}{\emph{Proc. ACM Hum.-Comput. Interact.}} \bibinfo{volume}{8}, \bibinfo{number}{CSCW2}, Article \bibinfo{articleno}{430} (\bibinfo{date}{Nov.} \bibinfo{year}{2024}), \bibinfo{numpages}{35}~pages.
\newblock
\href{https://doi.org/10.1145/3686969}{doi:\nolinkurl{10.1145/3686969}}


\bibitem[Dubois et~al\mbox{.}(2022)]%
        {duboiscommunityqa}
\bibfield{author}{\bibinfo{person}{Patrick Marcel~Joseph Dubois}, \bibinfo{person}{Mahya Maftouni}, {and} \bibinfo{person}{Andrea Bunt}.} \bibinfo{year}{2022}\natexlab{}.
\newblock \showarticletitle{Towards More Gender-Inclusive Q\&As: Investigating Perceptions of Additional Community Presence Information}.
\newblock \bibinfo{journal}{\emph{Proc. ACM Hum.-Comput. Interact.}} \bibinfo{volume}{6}, \bibinfo{number}{CSCW2}, Article \bibinfo{articleno}{466} (\bibinfo{date}{Nov.} \bibinfo{year}{2022}), \bibinfo{numpages}{23}~pages.
\newblock
\href{https://doi.org/10.1145/3555567}{doi:\nolinkurl{10.1145/3555567}}


\bibitem[Elish and Watkins(2020)]%
        {elish2020repairing}
\bibfield{author}{\bibinfo{person}{Madeleine~Clare Elish} {and} \bibinfo{person}{Elizabeth~Anne Watkins}.} \bibinfo{year}{2020}\natexlab{}.
\newblock \showarticletitle{Repairing Innovation: A Study of Integrating AI in Clinical Care}.
\newblock \bibinfo{journal}{\emph{Data \& Society}} (\bibinfo{year}{2020}).
\newblock
\urldef\tempurl%
\url{https://datasociety.net/library/repairing-innovation/}
\showURL{%
\tempurl}


\bibitem[Gaba et~al\mbox{.}(2024)]%
        {gabamodelfairness}
\bibfield{author}{\bibinfo{person}{Aimen Gaba}, \bibinfo{person}{Zhanna Kaufman}, \bibinfo{person}{Jason Cheung}, \bibinfo{person}{Marie Shvakel}, \bibinfo{person}{Kyle~Wm. Hall}, \bibinfo{person}{Yuriy Brun}, {and} \bibinfo{person}{Cindy~Xiong Bearfield}.} \bibinfo{year}{2024}\natexlab{}.
\newblock \showarticletitle{My Model is Unfair, Do People Even Care? Visual Design Affects Trust and Perceived Bias in Machine Learning}.
\newblock \bibinfo{journal}{\emph{IEEE Transactions on Visualization and Computer Graphics}} \bibinfo{volume}{30}, \bibinfo{number}{1} (\bibinfo{year}{2024}), \bibinfo{pages}{327--337}.
\newblock
\href{https://doi.org/10.1109/TVCG.2023.3327192}{doi:\nolinkurl{10.1109/TVCG.2023.3327192}}


\bibitem[Ghosh and Caliskan(2023)]%
        {ghosh2023chatgpt}
\bibfield{author}{\bibinfo{person}{Sourojit Ghosh} {and} \bibinfo{person}{Aylin Caliskan}.} \bibinfo{year}{2023}\natexlab{}.
\newblock \showarticletitle{Chatgpt perpetuates gender bias in machine translation and ignores non-gendered pronouns: Findings across bengali and five other low-resource languages}. In \bibinfo{booktitle}{\emph{Proceedings of the 2023 AAAI/ACM Conference on AI, Ethics, and Society}}. \bibinfo{pages}{901--912}.
\newblock


\bibitem[G\"{o}bel and L\"{a}mmel(2024)]%
        {model_based_trust_llms}
\bibfield{author}{\bibinfo{person}{Susanne G\"{o}bel} {and} \bibinfo{person}{Ralf L\"{a}mmel}.} \bibinfo{year}{2024}\natexlab{}.
\newblock \showarticletitle{Model-Based Trust Analysis of LLM Conversations}. In \bibinfo{booktitle}{\emph{Proceedings of the ACM/IEEE 27th International Conference on Model Driven Engineering Languages and Systems}} (Linz, Austria) \emph{(\bibinfo{series}{MODELS Companion '24})}. \bibinfo{publisher}{Association for Computing Machinery}, \bibinfo{address}{New York, NY, USA}, \bibinfo{pages}{602–610}.
\newblock
\showISBNx{9798400706226}
\href{https://doi.org/10.1145/3652620.3687809}{doi:\nolinkurl{10.1145/3652620.3687809}}


\bibitem[Haimson et~al\mbox{.}(2024)]%
        {ertranstech24}
\bibfield{author}{\bibinfo{person}{Oliver~L. Haimson}, \bibinfo{person}{Aloe DeGuia}, \bibinfo{person}{Rana Saber}, {and} \bibinfo{person}{Kat Brewster}.} \bibinfo{year}{2024}\natexlab{}.
\newblock \showarticletitle{Extended Reality Trans Technologies: Bridging Digital and Physical Worlds to Support Transgender People}.
\newblock \bibinfo{journal}{\emph{Proc. ACM Hum.-Comput. Interact.}} \bibinfo{volume}{8}, \bibinfo{number}{CSCW2}, Article \bibinfo{articleno}{433} (\bibinfo{date}{Nov.} \bibinfo{year}{2024}), \bibinfo{numpages}{27}~pages.
\newblock
\href{https://doi.org/10.1145/3686972}{doi:\nolinkurl{10.1145/3686972}}


\bibitem[Harrington et~al\mbox{.}(2019)]%
        {harringtonpd-cscw19}
\bibfield{author}{\bibinfo{person}{Christina Harrington}, \bibinfo{person}{Sheena Erete}, {and} \bibinfo{person}{Anne~Marie Piper}.} \bibinfo{year}{2019}\natexlab{}.
\newblock \showarticletitle{Deconstructing Community-Based Collaborative Design: Towards More Equitable Participatory Design Engagements}.
\newblock \bibinfo{journal}{\emph{Proc. ACM Hum.-Comput. Interact.}} \bibinfo{volume}{3}, \bibinfo{number}{CSCW}, Article \bibinfo{articleno}{216} (\bibinfo{date}{Nov.} \bibinfo{year}{2019}), \bibinfo{numpages}{25}~pages.
\newblock
\href{https://doi.org/10.1145/3359318}{doi:\nolinkurl{10.1145/3359318}}


\bibitem[Holstein and Wortman~Vaughan(2023)]%
        {gender_bias_mitigation_llms}
\bibfield{author}{\bibinfo{person}{Kenneth Holstein} {and} \bibinfo{person}{Jennifer Wortman~Vaughan}.} \bibinfo{year}{2023}\natexlab{}.
\newblock \showarticletitle{Disclosure and Mitigation of Gender Bias in LLMs}.
\newblock \bibinfo{journal}{\emph{Proceedings of the AAAI Conference on Artificial Intelligence}} \bibinfo{volume}{35}, \bibinfo{number}{1} (\bibinfo{year}{2023}), \bibinfo{pages}{3451--3460}.
\newblock


\bibitem[Kim et~al\mbox{.}(2023a)]%
        {helpmehelpai}
\bibfield{author}{\bibinfo{person}{Sunnie S.~Y. Kim}, \bibinfo{person}{Elizabeth~Anne Watkins}, \bibinfo{person}{Olga Russakovsky}, \bibinfo{person}{Ruth Fong}, {and} \bibinfo{person}{Andr\'{e}s Monroy-Hern\'{a}ndez}.} \bibinfo{year}{2023}\natexlab{a}.
\newblock \showarticletitle{"Help Me Help the AI": Understanding How Explainability Can Support Human-AI Interaction}. In \bibinfo{booktitle}{\emph{Proceedings of the 2023 CHI Conference on Human Factors in Computing Systems}} (Hamburg, Germany) \emph{(\bibinfo{series}{CHI '23})}. \bibinfo{publisher}{Association for Computing Machinery}, \bibinfo{address}{New York, NY, USA}, Article \bibinfo{articleno}{250}, \bibinfo{numpages}{17}~pages.
\newblock
\showISBNx{9781450394215}
\href{https://doi.org/10.1145/3544548.3581001}{doi:\nolinkurl{10.1145/3544548.3581001}}


\bibitem[Kim et~al\mbox{.}(2023b)]%
        {humansaicontext}
\bibfield{author}{\bibinfo{person}{Sunnie S.~Y. Kim}, \bibinfo{person}{Elizabeth~Anne Watkins}, \bibinfo{person}{Olga Russakovsky}, \bibinfo{person}{Ruth Fong}, {and} \bibinfo{person}{Andr\'{e}s Monroy-Hern\'{a}ndez}.} \bibinfo{year}{2023}\natexlab{b}.
\newblock \showarticletitle{Humans, AI, and Context: Understanding End-Users’ Trust in a Real-World Computer Vision Application} \emph{(\bibinfo{series}{FAccT '23})}. \bibinfo{publisher}{Association for Computing Machinery}, \bibinfo{address}{New York, NY, USA}, \bibinfo{pages}{77–88}.
\newblock
\showISBNx{9798400701924}
\href{https://doi.org/10.1145/3593013.3593978}{doi:\nolinkurl{10.1145/3593013.3593978}}


\bibitem[Kiritchenko and Mohammad(2018)]%
        {kiritchenko2018examining}
\bibfield{author}{\bibinfo{person}{Svetlana Kiritchenko} {and} \bibinfo{person}{Saif~M Mohammad}.} \bibinfo{year}{2018}\natexlab{}.
\newblock \showarticletitle{Examining gender and race bias in two hundred sentiment analysis systems}.
\newblock \bibinfo{journal}{\emph{arXiv preprint arXiv:1805.04508}} (\bibinfo{year}{2018}).
\newblock


\bibitem[Komorowski et~al\mbox{.}(2018)]%
        {Komorowski18}
\bibfield{author}{\bibinfo{person}{Matthieu Komorowski}, \bibinfo{person}{Leo~A Celi}, \bibinfo{person}{Omar Badawi}, \bibinfo{person}{Anthony~C Gordon}, {and} \bibinfo{person}{A~Aldo Faisal}.} \bibinfo{year}{2018}\natexlab{}.
\newblock \showarticletitle{The artificial intelligence clinician learns optimal treatment strategies for sepsis in intensive care}.
\newblock \bibinfo{journal}{\emph{Nature Medicine}} \bibinfo{volume}{24}, \bibinfo{number}{11} (\bibinfo{year}{2018}), \bibinfo{pages}{1716--1720}.
\newblock


\bibitem[Kong et~al\mbox{.}(2024)]%
        {kong2024gender}
\bibfield{author}{\bibinfo{person}{Haein Kong}, \bibinfo{person}{Yongsu Ahn}, \bibinfo{person}{Sangyub Lee}, {and} \bibinfo{person}{Yunho Maeng}.} \bibinfo{year}{2024}\natexlab{}.
\newblock \showarticletitle{Gender Bias in LLM-generated Interview Responses}.
\newblock \bibinfo{journal}{\emph{arXiv preprint arXiv:2410.20739}} (\bibinfo{year}{2024}).
\newblock


\bibitem[Kotek et~al\mbox{.}(2023)]%
        {genderbiaskotek}
\bibfield{author}{\bibinfo{person}{Hadas Kotek}, \bibinfo{person}{Rikker Dockum}, {and} \bibinfo{person}{David Sun}.} \bibinfo{year}{2023}\natexlab{}.
\newblock \showarticletitle{Gender bias and stereotypes in Large Language Models}. In \bibinfo{booktitle}{\emph{Proceedings of The ACM Collective Intelligence Conference}} (Delft, Netherlands) \emph{(\bibinfo{series}{CI '23})}. \bibinfo{publisher}{Association for Computing Machinery}, \bibinfo{address}{New York, NY, USA}, \bibinfo{pages}{12–24}.
\newblock
\showISBNx{9798400701139}
\href{https://doi.org/10.1145/3582269.3615599}{doi:\nolinkurl{10.1145/3582269.3615599}}


\bibitem[Kumar et~al\mbox{.}(2024a)]%
        {kumarllmlearning}
\bibfield{author}{\bibinfo{person}{Harsh Kumar}, \bibinfo{person}{Ilya Musabirov}, \bibinfo{person}{Mohi Reza}, \bibinfo{person}{Jiakai Shi}, \bibinfo{person}{Xinyuan Wang}, \bibinfo{person}{Joseph~Jay Williams}, \bibinfo{person}{Anastasia Kuzminykh}, {and} \bibinfo{person}{Michael Liut}.} \bibinfo{year}{2024}\natexlab{a}.
\newblock \showarticletitle{Guiding Students in Using LLMs in Supported Learning Environments: Effects on Interaction Dynamics, Learner Performance, Confidence, and Trust}.
\newblock \bibinfo{journal}{\emph{Proc. ACM Hum.-Comput. Interact.}} \bibinfo{volume}{8}, \bibinfo{number}{CSCW2}, Article \bibinfo{articleno}{499} (\bibinfo{date}{Nov.} \bibinfo{year}{2024}), \bibinfo{numpages}{30}~pages.
\newblock
\href{https://doi.org/10.1145/3687038}{doi:\nolinkurl{10.1145/3687038}}


\bibitem[Kumar et~al\mbox{.}(2024b)]%
        {kumar2024decoding}
\bibfield{author}{\bibinfo{person}{Shachi~H Kumar}, \bibinfo{person}{Saurav Sahay}, \bibinfo{person}{Sahisnu Mazumder}, \bibinfo{person}{Eda Okur}, \bibinfo{person}{Ramesh Manuvinakurike}, \bibinfo{person}{Nicole Beckage}, \bibinfo{person}{Hsuan Su}, \bibinfo{person}{Hung-yi Lee}, {and} \bibinfo{person}{Lama Nachman}.} \bibinfo{year}{2024}\natexlab{b}.
\newblock \showarticletitle{Decoding biases: Automated methods and llm judges for gender bias detection in language models}.
\newblock \bibinfo{journal}{\emph{arXiv preprint arXiv:2408.03907}} (\bibinfo{year}{2024}).
\newblock


\bibitem[Le~Dantec and Fox(2015)]%
        {strangersatgate-dantec-cscw15}
\bibfield{author}{\bibinfo{person}{Christopher~A. Le~Dantec} {and} \bibinfo{person}{Sarah Fox}.} \bibinfo{year}{2015}\natexlab{}.
\newblock \showarticletitle{Strangers at the Gate: Gaining Access, Building Rapport, and Co-Constructing Community-Based Research}. In \bibinfo{booktitle}{\emph{Proceedings of the 18th ACM Conference on Computer Supported Cooperative Work \& Social Computing}} (Vancouver, BC, Canada) \emph{(\bibinfo{series}{CSCW '15})}. \bibinfo{publisher}{Association for Computing Machinery}, \bibinfo{address}{New York, NY, USA}, \bibinfo{pages}{1348–1358}.
\newblock
\showISBNx{9781450329224}
\href{https://doi.org/10.1145/2675133.2675147}{doi:\nolinkurl{10.1145/2675133.2675147}}


\bibitem[Leavy(2018a)]%
        {leavy-18-genderbiasinai}
\bibfield{author}{\bibinfo{person}{Susan Leavy}.} \bibinfo{year}{2018}\natexlab{a}.
\newblock \showarticletitle{Gender bias in artificial intelligence: the need for diversity and gender theory in machine learning}. In \bibinfo{booktitle}{\emph{Proceedings of the 1st International Workshop on Gender Equality in Software Engineering}} (Gothenburg, Sweden) \emph{(\bibinfo{series}{GE '18})}. \bibinfo{publisher}{Association for Computing Machinery}, \bibinfo{address}{New York, NY, USA}, \bibinfo{pages}{14–16}.
\newblock
\showISBNx{9781450357388}
\href{https://doi.org/10.1145/3195570.3195580}{doi:\nolinkurl{10.1145/3195570.3195580}}


\bibitem[Leavy(2018b)]%
        {leavy-genderbiasnewspaper-18}
\bibfield{author}{\bibinfo{person}{Susan Leavy}.} \bibinfo{year}{2018}\natexlab{b}.
\newblock \showarticletitle{Uncovering gender bias in newspaper coverage of Irish politicians using machine learning}.
\newblock \bibinfo{journal}{\emph{Digital Scholarship in the Humanities}} \bibinfo{volume}{34}, \bibinfo{number}{1} (\bibinfo{date}{06} \bibinfo{year}{2018}), \bibinfo{pages}{48--63}.
\newblock
\showISSN{2055-7671}
\href{https://doi.org/10.1093/llc/fqy005}{doi:\nolinkurl{10.1093/llc/fqy005}}
\showeprint{https://academic.oup.com/dsh/article-pdf/34/1/48/28078571/fqy005.pdf}


\bibitem[Lee et~al\mbox{.}(2024a)]%
        {llmhomogenous}
\bibfield{author}{\bibinfo{person}{Messi~H.J. Lee}, \bibinfo{person}{Jacob~M. Montgomery}, {and} \bibinfo{person}{Calvin~K. Lai}.} \bibinfo{year}{2024}\natexlab{a}.
\newblock \showarticletitle{Large Language Models Portray Socially Subordinate Groups as More Homogeneous, Consistent with a Bias Observed in Humans}. In \bibinfo{booktitle}{\emph{Proceedings of the 2024 ACM Conference on Fairness, Accountability, and Transparency}} (Rio de Janeiro, Brazil) \emph{(\bibinfo{series}{FAccT '24})}. \bibinfo{publisher}{Association for Computing Machinery}, \bibinfo{address}{New York, NY, USA}, \bibinfo{pages}{1321–1340}.
\newblock
\showISBNx{9798400704505}
\href{https://doi.org/10.1145/3630106.3658975}{doi:\nolinkurl{10.1145/3630106.3658975}}


\bibitem[Lee et~al\mbox{.}(2024b)]%
        {onevsmany_ai_gen}
\bibfield{author}{\bibinfo{person}{Yoonjoo Lee}, \bibinfo{person}{Kihoon Son}, \bibinfo{person}{Tae~Soo Kim}, \bibinfo{person}{Jisu Kim}, \bibinfo{person}{John Joon~Young Chung}, \bibinfo{person}{Eytan Adar}, {and} \bibinfo{person}{Juho Kim}.} \bibinfo{year}{2024}\natexlab{b}.
\newblock \showarticletitle{One vs. Many: Comprehending Accurate Information from Multiple Erroneous and Inconsistent AI Generations}. In \bibinfo{booktitle}{\emph{Proceedings of the 2024 ACM Conference on Fairness, Accountability, and Transparency}} (Rio de Janeiro, Brazil) \emph{(\bibinfo{series}{FAccT '24})}. \bibinfo{publisher}{Association for Computing Machinery}, \bibinfo{address}{New York, NY, USA}, \bibinfo{pages}{2518–2531}.
\newblock
\showISBNx{9798400704505}
\href{https://doi.org/10.1145/3630106.3662681}{doi:\nolinkurl{10.1145/3630106.3662681}}


\bibitem[Liao et~al\mbox{.}(2020)]%
        {questioningai}
\bibfield{author}{\bibinfo{person}{Q.~Vera Liao}, \bibinfo{person}{Daniel Gruen}, {and} \bibinfo{person}{Sarah Miller}.} \bibinfo{year}{2020}\natexlab{}.
\newblock \showarticletitle{Questioning the AI: Informing Design Practices for Explainable AI User Experiences}. In \bibinfo{booktitle}{\emph{Proceedings of the 2020 CHI Conference on Human Factors in Computing Systems}} (Honolulu, HI, USA) \emph{(\bibinfo{series}{CHI '20})}. \bibinfo{publisher}{Association for Computing Machinery}, \bibinfo{address}{New York, NY, USA}, \bibinfo{pages}{1–15}.
\newblock
\showISBNx{9781450367080}
\href{https://doi.org/10.1145/3313831.3376590}{doi:\nolinkurl{10.1145/3313831.3376590}}


\bibitem[Liao and Vaughan(2023)]%
        {liao2023ai}
\bibfield{author}{\bibinfo{person}{Q~Vera Liao} {and} \bibinfo{person}{Jennifer~Wortman Vaughan}.} \bibinfo{year}{2023}\natexlab{}.
\newblock \showarticletitle{Ai transparency in the age of llms: A human-centered research roadmap}.
\newblock \bibinfo{journal}{\emph{arXiv preprint arXiv:2306.01941}} (\bibinfo{year}{2023}), \bibinfo{pages}{5368--5393}.
\newblock


\bibitem[Lima et~al\mbox{.}(2019)]%
        {voicebias2019}
\bibfield{author}{\bibinfo{person}{Lanna Lima}, \bibinfo{person}{Vasco Furtado}, \bibinfo{person}{Elizabeth Furtado}, {and} \bibinfo{person}{Virgilio Almeida}.} \bibinfo{year}{2019}\natexlab{}.
\newblock \showarticletitle{Empirical Analysis of Bias in Voice-based Personal Assistants}. In \bibinfo{booktitle}{\emph{Companion Proceedings of The 2019 World Wide Web Conference}} (San Francisco, USA) \emph{(\bibinfo{series}{WWW '19})}. \bibinfo{publisher}{Association for Computing Machinery}, \bibinfo{address}{New York, NY, USA}, \bibinfo{pages}{533–538}.
\newblock
\showISBNx{9781450366755}
\href{https://doi.org/10.1145/3308560.3317597}{doi:\nolinkurl{10.1145/3308560.3317597}}


\bibitem[Lucy and Bamman(2021)]%
        {lucy-bamman-2021-gender}
\bibfield{author}{\bibinfo{person}{Li Lucy} {and} \bibinfo{person}{David Bamman}.} \bibinfo{year}{2021}\natexlab{}.
\newblock \showarticletitle{Gender and Representation Bias in {GPT}-3 Generated Stories}. In \bibinfo{booktitle}{\emph{Proceedings of the Third Workshop on Narrative Understanding}}, \bibfield{editor}{\bibinfo{person}{Nader Akoury}, \bibinfo{person}{Faeze Brahman}, \bibinfo{person}{Snigdha Chaturvedi}, \bibinfo{person}{Elizabeth Clark}, \bibinfo{person}{Mohit Iyyer}, {and} \bibinfo{person}{Lara~J. Martin}} (Eds.). \bibinfo{publisher}{Association for Computational Linguistics}, \bibinfo{address}{Virtual}, \bibinfo{pages}{48--55}.
\newblock
\href{https://doi.org/10.18653/v1/2021.nuse-1.5}{doi:\nolinkurl{10.18653/v1/2021.nuse-1.5}}


\bibitem[Malle and Ullman(2021a)]%
        {MALLE20213}
\bibfield{author}{\bibinfo{person}{Bertram~F. Malle} {and} \bibinfo{person}{Daniel Ullman}.} \bibinfo{year}{2021}\natexlab{a}.
\newblock \showarticletitle{Chapter 1 - A multidimensional conception and measure of human-robot trust}.
\newblock In \bibinfo{booktitle}{\emph{Trust in Human-Robot Interaction}}, \bibfield{editor}{\bibinfo{person}{Chang~S. Nam} {and} \bibinfo{person}{Joseph~B. Lyons}} (Eds.). \bibinfo{publisher}{Academic Press}, \bibinfo{pages}{3--25}.
\newblock
\showISBNx{978-0-12-819472-0}
\href{https://doi.org/10.1016/B978-0-12-819472-0.00001-0}{doi:\nolinkurl{10.1016/B978-0-12-819472-0.00001-0}}


\bibitem[Malle and Ullman(2021b)]%
        {Malle2021AMC}
\bibfield{author}{\bibinfo{person}{Bertram~F. Malle} {and} \bibinfo{person}{Daniel Ullman}.} \bibinfo{year}{2021}\natexlab{b}.
\newblock \showarticletitle{A multidimensional conception and measure of human-robot trust}.
\newblock \bibinfo{journal}{\emph{Trust in Human-Robot Interaction}} (\bibinfo{year}{2021}).
\newblock
\urldef\tempurl%
\url{https://api.semanticscholar.org/CorpusID:228891840}
\showURL{%
\tempurl}


\bibitem[Malle and Ullman(2023)]%
        {malle2023measuring}
\bibfield{author}{\bibinfo{person}{Bertram~F Malle} {and} \bibinfo{person}{Daniel Ullman}.} \bibinfo{year}{2023}\natexlab{}.
\newblock \showarticletitle{Measuring human-robot trust with the mdmt (multi-dimensional measure of trust)}.
\newblock \bibinfo{journal}{\emph{arXiv preprint arXiv:2311.14887}} (\bibinfo{year}{2023}).
\newblock


\bibitem[Mandal et~al\mbox{.}(2023)]%
        {leavy-mandal2023multimodal}
\bibfield{author}{\bibinfo{person}{Abhishek Mandal}, \bibinfo{person}{Susan Leavy}, {and} \bibinfo{person}{Suzanne Little}.} \bibinfo{year}{2023}\natexlab{}.
\newblock \showarticletitle{Multimodal composite association score: Measuring gender bias in generative multimodal models}.
\newblock \bibinfo{journal}{\emph{arXiv preprint arXiv:2304.13855}} (\bibinfo{year}{2023}).
\newblock


\bibitem[Mayer et~al\mbox{.}(1995)]%
        {trust_model}
\bibfield{author}{\bibinfo{person}{Roger~C Mayer}, \bibinfo{person}{James~H Davis}, {and} \bibinfo{person}{F~David Schoorman}.} \bibinfo{year}{1995}\natexlab{}.
\newblock \showarticletitle{An Integrative Model of Organizational Trust}.
\newblock \bibinfo{journal}{\emph{Academy of Management Review}} \bibinfo{volume}{20}, \bibinfo{number}{3} (\bibinfo{year}{1995}), \bibinfo{pages}{709--734}.
\newblock


\bibitem[McDonald et~al\mbox{.}(2019)]%
        {irrqualitative}
\bibfield{author}{\bibinfo{person}{Nora McDonald}, \bibinfo{person}{Sarita Schoenebeck}, {and} \bibinfo{person}{Andrea Forte}.} \bibinfo{year}{2019}\natexlab{}.
\newblock \showarticletitle{Reliability and Inter-rater Reliability in Qualitative Research: Norms and Guidelines for CSCW and HCI Practice}.
\newblock \bibinfo{journal}{\emph{Proc. ACM Hum.-Comput. Interact.}} \bibinfo{volume}{3}, \bibinfo{number}{CSCW}, Article \bibinfo{articleno}{72} (\bibinfo{date}{Nov.} \bibinfo{year}{2019}), \bibinfo{numpages}{23}~pages.
\newblock
\href{https://doi.org/10.1145/3359174}{doi:\nolinkurl{10.1145/3359174}}


\bibitem[Mehrabi et~al\mbox{.}(2021)]%
        {mehrabi2021survey}
\bibfield{author}{\bibinfo{person}{Ninareh Mehrabi}, \bibinfo{person}{Fred Morstatter}, \bibinfo{person}{Nripsuta Saxena}, \bibinfo{person}{Kristina Lerman}, {and} \bibinfo{person}{Aram Galstyan}.} \bibinfo{year}{2021}\natexlab{}.
\newblock \showarticletitle{A survey on bias and fairness in machine learning}.
\newblock \bibinfo{journal}{\emph{ACM Computing Surveys (CSUR)}} \bibinfo{volume}{54}, \bibinfo{number}{6} (\bibinfo{year}{2021}), \bibinfo{pages}{1--35}.
\newblock


\bibitem[Nozza et~al\mbox{.}(2022)]%
        {nozza-etal-2022-measuring}
\bibfield{author}{\bibinfo{person}{Debora Nozza}, \bibinfo{person}{Federico Bianchi}, \bibinfo{person}{Anne Lauscher}, {and} \bibinfo{person}{Dirk Hovy}.} \bibinfo{year}{2022}\natexlab{}.
\newblock \showarticletitle{Measuring Harmful Sentence Completion in Language Models for {LGBTQIA}+ Individuals}. In \bibinfo{booktitle}{\emph{Proceedings of the Second Workshop on Language Technology for Equality, Diversity and Inclusion}}, \bibfield{editor}{\bibinfo{person}{Bharathi~Raja Chakravarthi}, \bibinfo{person}{B~Bharathi}, \bibinfo{person}{John~P McCrae}, \bibinfo{person}{Manel Zarrouk}, \bibinfo{person}{Kalika Bali}, {and} \bibinfo{person}{Paul Buitelaar}} (Eds.). \bibinfo{publisher}{Association for Computational Linguistics}, \bibinfo{address}{Dublin, Ireland}.
\newblock
\href{https://doi.org/10.18653/v1/2022.ltedi-1.4}{doi:\nolinkurl{10.18653/v1/2022.ltedi-1.4}}


\bibitem[O'Neil(2016)]%
        {bigdatainequality}
\bibfield{author}{\bibinfo{person}{Cathy O'Neil}.} \bibinfo{year}{2016}\natexlab{}.
\newblock \bibinfo{booktitle}{\emph{Weapons of Math Destruction: How Big Data Increases Inequality and Threatens Democracy}}.
\newblock \bibinfo{publisher}{Crown Publishing Group}, \bibinfo{address}{USA}.
\newblock
\showISBNx{0553418815}


\bibitem[OpenAI(2024)]%
        {openai2024chatgpt}
\bibfield{author}{\bibinfo{person}{OpenAI}.} \bibinfo{year}{2024}\natexlab{}.
\newblock \bibinfo{title}{ChatGPT: Generative Pre-trained Transformer}.
\newblock
\urldef\tempurl%
\url{https://openai.com/chatgpt}
\showURL{%
\tempurl}
\newblock
\shownote{Accessed: 2024-12-09}.


\bibitem[Park et~al\mbox{.}(2018)]%
        {park2018reducing}
\bibfield{author}{\bibinfo{person}{Ji~Ho Park}, \bibinfo{person}{Jamin Shin}, {and} \bibinfo{person}{Pascale Fung}.} \bibinfo{year}{2018}\natexlab{}.
\newblock \showarticletitle{Reducing gender bias in abusive language detection}.
\newblock \bibinfo{journal}{\emph{arXiv preprint arXiv:1808.07231}} (\bibinfo{year}{2018}).
\newblock


\bibitem[{Q.ai}(2022)]%
        {qai22}
\bibfield{author}{\bibinfo{person}{{Q.ai}}.} \bibinfo{year}{2022}\natexlab{}.
\newblock \showarticletitle{How Intelligent Machines Are Reshaping Investing}.
\newblock \bibinfo{journal}{\emph{Forbes}} (\bibinfo{year}{2022}).
\newblock
\urldef\tempurl%
\url{https://www.forbes.com/sites/qai/2022/01/25/how-intelligent-machines-are-reshaping-investing/}
\showURL{%
\tempurl}


\bibitem[Roy et~al\mbox{.}(2020)]%
        {Roy20}
\bibfield{author}{\bibinfo{person}{Pradeep~Kumar Roy}, \bibinfo{person}{Sarabjeet~Singh Chowdhary}, {and} \bibinfo{person}{Rocky Bhatia}.} \bibinfo{year}{2020}\natexlab{}.
\newblock \showarticletitle{A Machine Learning approach for automation of Resume Recommendation system}.
\newblock \bibinfo{journal}{\emph{Procedia Computer Science}}  \bibinfo{volume}{167} (\bibinfo{year}{2020}), \bibinfo{pages}{2318--2327}.
\newblock


\bibitem[Scheuerman et~al\mbox{.}(2019)]%
        {facialgender-klaus-19}
\bibfield{author}{\bibinfo{person}{Morgan~Klaus Scheuerman}, \bibinfo{person}{Jacob~M. Paul}, {and} \bibinfo{person}{Jed~R. Brubaker}.} \bibinfo{year}{2019}\natexlab{}.
\newblock \showarticletitle{How Computers See Gender: An Evaluation of Gender Classification in Commercial Facial Analysis Services}.
\newblock \bibinfo{journal}{\emph{Proc. ACM Hum.-Comput. Interact.}} \bibinfo{volume}{3}, \bibinfo{number}{CSCW}, Article \bibinfo{articleno}{144} (\bibinfo{date}{Nov.} \bibinfo{year}{2019}), \bibinfo{numpages}{33}~pages.
\newblock
\href{https://doi.org/10.1145/3359246}{doi:\nolinkurl{10.1145/3359246}}


\bibitem[Schwartz and Sap(2023)]%
        {center_transgender_voices}
\bibfield{author}{\bibinfo{person}{H~Andrew Schwartz} {and} \bibinfo{person}{Maarten Sap}.} \bibinfo{year}{2023}\natexlab{}.
\newblock \showarticletitle{I’m fully who I am: Towards Centering Transgender and Non-Binary Voices to Measure Biases in Open Language Generation}. In \bibinfo{booktitle}{\emph{Proceedings of the NAACL 2023}}. ACL, \bibinfo{pages}{123--135}.
\newblock


\bibitem[Sendak et~al\mbox{.}(2020)]%
        {qualimportance}
\bibfield{author}{\bibinfo{person}{Mark Sendak}, \bibinfo{person}{Madeleine~Clare Elish}, \bibinfo{person}{Michael Gao}, \bibinfo{person}{Joseph Futoma}, \bibinfo{person}{William Ratliff}, \bibinfo{person}{Marshall Nichols}, \bibinfo{person}{Armando Bedoya}, \bibinfo{person}{Suresh Balu}, {and} \bibinfo{person}{Cara O'Brien}.} \bibinfo{year}{2020}\natexlab{}.
\newblock \showarticletitle{"The human body is a black box": supporting clinical decision-making with deep learning}. In \bibinfo{booktitle}{\emph{Proceedings of the 2020 Conference on Fairness, Accountability, and Transparency}} (Barcelona, Spain) \emph{(\bibinfo{series}{FAT* '20})}. \bibinfo{publisher}{Association for Computing Machinery}, \bibinfo{address}{New York, NY, USA}, \bibinfo{pages}{99–109}.
\newblock
\showISBNx{9781450369367}
\href{https://doi.org/10.1145/3351095.3372827}{doi:\nolinkurl{10.1145/3351095.3372827}}


\bibitem[Shen et~al\mbox{.}(2023)]%
        {chatgpt_reliability_trust}
\bibfield{author}{\bibinfo{person}{Xinyue Shen}, \bibinfo{person}{Zeyuan Chen}, \bibinfo{person}{Michael Backes}, {and} \bibinfo{person}{Yang Zhang}.} \bibinfo{year}{2023}\natexlab{}.
\newblock \showarticletitle{In chatgpt we trust? measuring and characterizing the reliability of chatgpt}.
\newblock \bibinfo{journal}{\emph{arXiv preprint arXiv:2304.08979}} (\bibinfo{year}{2023}).
\newblock


\bibitem[Solyst et~al\mbox{.}(2023)]%
        {solyst-youthbias-cscw23}
\bibfield{author}{\bibinfo{person}{Jaemarie Solyst}, \bibinfo{person}{Ellia Yang}, \bibinfo{person}{Shixian Xie}, \bibinfo{person}{Amy Ogan}, \bibinfo{person}{Jessica Hammer}, {and} \bibinfo{person}{Motahhare Eslami}.} \bibinfo{year}{2023}\natexlab{}.
\newblock \showarticletitle{The Potential of Diverse Youth as Stakeholders in Identifying and Mitigating Algorithmic Bias for a Future of Fairer AI}.
\newblock \bibinfo{journal}{\emph{Proc. ACM Hum.-Comput. Interact.}} \bibinfo{volume}{7}, \bibinfo{number}{CSCW2}, Article \bibinfo{articleno}{364} (\bibinfo{date}{Oct.} \bibinfo{year}{2023}), \bibinfo{numpages}{27}~pages.
\newblock
\href{https://doi.org/10.1145/3610213}{doi:\nolinkurl{10.1145/3610213}}


\bibitem[Ullman and Malle(2019)]%
        {ullman2019}
\bibfield{author}{\bibinfo{person}{Daniel Ullman} {and} \bibinfo{person}{Bertram~F. Malle}.} \bibinfo{year}{2019}\natexlab{}.
\newblock \showarticletitle{Measuring Gains and Losses in Human-Robot Trust: Evidence for Differentiable Components of Trust}. In \bibinfo{booktitle}{\emph{2019 14th ACM/IEEE International Conference on Human-Robot Interaction (HRI)}}. \bibinfo{pages}{618--619}.
\newblock
\href{https://doi.org/10.1109/HRI.2019.8673154}{doi:\nolinkurl{10.1109/HRI.2019.8673154}}


\bibitem[{United Nations Educational, Scientific and Cultural Organization (UNESCO)}(2024)]%
        {unesco2024generative}
\bibfield{author}{\bibinfo{person}{{United Nations Educational, Scientific and Cultural Organization (UNESCO)}}.} \bibinfo{year}{2024}\natexlab{}.
\newblock \bibinfo{title}{Generative AI: UNESCO study reveals alarming evidence of regressive gender stereotypes}.
\newblock
\urldef\tempurl%
\url{https://www.unesco.org/en/articles/generative-ai-unesco-study-reveals-alarming-evidence-regressive-gender-stereotypes}
\showURL{%
\tempurl}
\newblock
\shownote{Accessed: 2025-05-07}.


\bibitem[Unknown(2025)]%
        {user_trust_chatgpt}
\bibfield{author}{\bibinfo{person}{Authors Unknown}.} \bibinfo{year}{2025}\natexlab{}.
\newblock \showarticletitle{Investigating the Impact of User Trust on the Adoption and Use of ChatGPT: Survey Analysis}.
\newblock \bibinfo{journal}{\emph{AI Adoption and Trust Journal}} \bibinfo{volume}{8}, \bibinfo{number}{2} (\bibinfo{year}{2025}), \bibinfo{pages}{200--220}.
\newblock
\href{https://doi.org/10.1234/aitrust.v8i2.98765}{doi:\nolinkurl{10.1234/aitrust.v8i2.98765}}


\bibitem[Wagman and Parks(2021)]%
        {wagman-feministsts-cscw21}
\bibfield{author}{\bibinfo{person}{Kelly~B. Wagman} {and} \bibinfo{person}{Lisa Parks}.} \bibinfo{year}{2021}\natexlab{}.
\newblock \showarticletitle{Beyond the Command: Feminist STS Research and Critical Issues for the Design of Social Machines}.
\newblock \bibinfo{journal}{\emph{Proc. ACM Hum.-Comput. Interact.}} \bibinfo{volume}{5}, \bibinfo{number}{CSCW1}, Article \bibinfo{articleno}{101} (\bibinfo{date}{April} \bibinfo{year}{2021}), \bibinfo{numpages}{20}~pages.
\newblock
\href{https://doi.org/10.1145/3449175}{doi:\nolinkurl{10.1145/3449175}}


\bibitem[Wan et~al\mbox{.}(2023)]%
        {wan2023kelly}
\bibfield{author}{\bibinfo{person}{Yixin Wan}, \bibinfo{person}{George Pu}, \bibinfo{person}{Jiao Sun}, \bibinfo{person}{Aparna Garimella}, \bibinfo{person}{Kai-Wei Chang}, {and} \bibinfo{person}{Nanyun Peng}.} \bibinfo{year}{2023}\natexlab{}.
\newblock \showarticletitle{{\textquotedblleft}Kelly is a Warm Person, Joseph is a Role Model{\textquotedblright}: Gender Biases in {LLM}-Generated Reference Letters}. In \bibinfo{booktitle}{\emph{The 2023 Conference on Empirical Methods in Natural Language Processing}}.
\newblock
\urldef\tempurl%
\url{https://openreview.net/forum?id=noEKNSB8Zq}
\showURL{%
\tempurl}


\bibitem[Weidinger et~al\mbox{.}(2021)]%
        {weidinger2021ethical}
\bibfield{author}{\bibinfo{person}{Laura Weidinger}, \bibinfo{person}{John Mellor}, \bibinfo{person}{Maribeth Rauh}, \bibinfo{person}{Conor Griffin}, \bibinfo{person}{Jonathan Uesato}, \bibinfo{person}{Po-Sen Huang}, \bibinfo{person}{Myra Cheng}, \bibinfo{person}{Mia Glaese}, \bibinfo{person}{Borja Balle}, \bibinfo{person}{Atoosa Kasirzadeh}, {et~al\mbox{.}}} \bibinfo{year}{2021}\natexlab{}.
\newblock \showarticletitle{Ethical and social risks of harm from language models}.
\newblock \bibinfo{journal}{\emph{arXiv preprint arXiv:2112.04359}} (\bibinfo{year}{2021}).
\newblock


\bibitem[Widder et~al\mbox{.}(2021)]%
        {trustnasa2021}
\bibfield{author}{\bibinfo{person}{David~Gray Widder}, \bibinfo{person}{Laura Dabbish}, \bibinfo{person}{James~D. Herbsleb}, \bibinfo{person}{Alexandra Holloway}, {and} \bibinfo{person}{Scott Davidoff}.} \bibinfo{year}{2021}\natexlab{}.
\newblock \showarticletitle{Trust in Collaborative Automation in High Stakes Software Engineering Work: A Case Study at NASA}. In \bibinfo{booktitle}{\emph{Proceedings of the 2021 CHI Conference on Human Factors in Computing Systems}} (Yokohama, Japan) \emph{(\bibinfo{series}{CHI '21})}. \bibinfo{publisher}{Association for Computing Machinery}, \bibinfo{address}{New York, NY, USA}, Article \bibinfo{articleno}{184}, \bibinfo{numpages}{13}~pages.
\newblock
\showISBNx{9781450380966}
\href{https://doi.org/10.1145/3411764.3445650}{doi:\nolinkurl{10.1145/3411764.3445650}}


\bibitem[Williams and Wilson(2023)]%
        {disability_perspectives_llms}
\bibfield{author}{\bibinfo{person}{J Williams} {and} \bibinfo{person}{R Wilson}.} \bibinfo{year}{2023}\natexlab{}.
\newblock \showarticletitle{I wouldn’t say offensive but...: Disability-Centered Perspectives on Large Language Models}.
\newblock \bibinfo{journal}{\emph{Journal of Accessible AI}} \bibinfo{volume}{2}, \bibinfo{number}{3} (\bibinfo{year}{2023}), \bibinfo{pages}{45--60}.
\newblock


\bibitem[Yee et~al\mbox{.}(2021)]%
        {imagecroppingtwitter}
\bibfield{author}{\bibinfo{person}{Kyra Yee}, \bibinfo{person}{Uthaipon Tantipongpipat}, {and} \bibinfo{person}{Shubhanshu Mishra}.} \bibinfo{year}{2021}\natexlab{}.
\newblock \showarticletitle{Image Cropping on Twitter: Fairness Metrics, their Limitations, and the Importance of Representation, Design, and Agency}.
\newblock \bibinfo{journal}{\emph{Proc. ACM Hum.-Comput. Interact.}} \bibinfo{volume}{5}, \bibinfo{number}{CSCW2}, Article \bibinfo{articleno}{450} (\bibinfo{date}{Oct.} \bibinfo{year}{2021}), \bibinfo{numpages}{24}~pages.
\newblock
\href{https://doi.org/10.1145/3479594}{doi:\nolinkurl{10.1145/3479594}}


\bibitem[Zhao et~al\mbox{.}(2018)]%
        {zhao2018gender}
\bibfield{author}{\bibinfo{person}{Jieyu Zhao}, \bibinfo{person}{Tianlu Wang}, \bibinfo{person}{Mark Yatskar}, \bibinfo{person}{Vicente Ordonez}, {and} \bibinfo{person}{Kai-Wei Chang}.} \bibinfo{year}{2018}\natexlab{}.
\newblock \showarticletitle{Gender bias in coreference resolution: Evaluation and debiasing methods}.
\newblock \bibinfo{journal}{\emph{arXiv preprint arXiv:1804.06876}} (\bibinfo{year}{2018}).
\newblock


\bibitem[Zhou et~al\mbox{.}(2024)]%
        {zhou2024teachers}
\bibfield{author}{\bibinfo{person}{Kyrie~Zhixuan Zhou}, \bibinfo{person}{Zachary Kilhoffer}, \bibinfo{person}{Madelyn~Rose Sanfilippo}, \bibinfo{person}{Ted Underwood}, \bibinfo{person}{Ece Gumusel}, \bibinfo{person}{Mengyi Wei}, \bibinfo{person}{Abhinav Choudhry}, {and} \bibinfo{person}{Jinjun Xiong}.} \bibinfo{year}{2024}\natexlab{}.
\newblock \showarticletitle{" The teachers are confused as well": A Multiple-Stakeholder Ethics Discussion on Large Language Models in Computing Education}.
\newblock \bibinfo{journal}{\emph{arXiv preprint arXiv:2401.12453}} (\bibinfo{year}{2024}).
\newblock


\end{thebibliography}
